\documentclass[a4paper,11pt]{article}
\pdfoutput=1 

\usepackage{jheppub} 

\usepackage[T1]{fontenc} 
\usepackage{amsmath}
\usepackage{amssymb}
\usepackage{tikz}
\usepackage{tikz-feynman}
\usetikzlibrary{feynman}
\usetikzlibrary{positioning}
\allowdisplaybreaks
\title{Scalar quasinormal modes of Kerr--AdS$_{\bf 7}$ via accessory parameter expansions}

\author[a]{Juli\'{a}n Barrag\'{a}n Amado}

\affiliation[a]{Grupo de F\'{i}sica Matem\'{a}tica,
Departamento de Matem\'{a}tica, Instituto Superior T\'{e}cnico -- Universidade de Lisboa, Avenida Rovisco Pais 1, 1049-001 Lisboa, Portugal}

\emailAdd{jose.barragan.amado@tecnico.ulisboa.pt}

\abstract{We study scalar perturbations of a seven-dimensional Kerr--AdS black hole with three independent rotation parameters. In this geometry, the Klein--Gordon equation separates into one radial and two angular second-order linear ordinary differential equations, each with five regular singular points. We apply the Hill determinant method to compute analytic expansions of the accessory parameters in two different parameter regimes. The resulting accessory parameter expansions coincide with those obtained from the instanton part of the Nekrasov--Shatashvili function of four-dimensional $\mathcal{N}=2$ quiver gauge theories. These expansions are then used to derive perturbative expressions for the angular eigenvalues in the slowly rotating limit and for the quasinormal mode frequencies in the small black hole limit.}

\begin{document} 
\maketitle
\flushbottom

\section{Introduction}
\label{sec:1}

Black holes in higher dimensions are of particular interest in supergravity and string theory~\cite{Emparan:2008eg}. Within the class of higher-dimensional black objects, rotating black holes exhibit two distinctive features that are intrinsically related to the spacetime dimension. First, rotation can occur independently in several orthogonal two-planes, each associated with an independent angular momentum. Second, there is a qualitative change in the competition between attractive gravitational potential and the centrifugal repulsion barrier. Specifically, the radial fall-off of the Newtonian potential depends on the number of dimensions, whereas the centrifugal barrier does not, since rotation is confined to a plane~\cite{Emparan:2003sy}.

The higher-dimensional generalization of the Kerr geometry was obtained by Myers and Perry~\cite{Myers:1986un}. A first generalization of the Myers--Perry solution that includes the cosmological constant was found in~\cite{Hawking:1998kw}, and was later extended to arbitrary dimensions in~\cite{Gibbons:2004uw,Gibbons:2004js}. Finally, the most general Kerr--NUT--(A)dS in $D$ dimensions was found by Chen, L\"{u}, and Pope~\cite{Chen:2006xh}. These higher-dimensional black hole solutions contain $\lfloor\frac{D-1}{2}\rfloor$ independent rotation parameters, corresponding to independent rotations in $\lfloor\frac{D-1}{2}\rfloor$ orthogonal spatial two-planes. Another remarkable property of the Kerr--NUT--(A)dS metrics is the existence of a rank-2 closed conformal Killing--Yano tensor, which generates the tower of Killing--Yano and Killing tensors that guarantee the full integrability of the geodesic equation~\cite{Page:2006ka}, as well as the separation of variables for the Hamilton--Jacobi, Klein--Gordon, and Dirac equations~\cite{Frolov:2006pe,Oota:2007vx}. The separability of the spin-1 field equation was demonstrated in~\cite{Lunin:2017drx,Frolov:2018ezx}. For reviews on hidden symmetries and complete integrability, see, for example,~\cite{Frolov:2008jr,Frolov:2017kze}.

After separation of variables, the equations of motion for field perturbations typically reduce to a set of second-order linear ordinary differential equations (ODEs) with a finite number of singular points, which can be written as Heun-type equations. Thus, higher-dimensional black holes can serve as probes of second-order linear ODEs and their associated connection problems, and vice versa. For instance, five-dimensional asymptotically AdS black holes have been used to study the accessory parameter problem of the Heun equation, which in turn allowed the computation of quasinormal modes of scalar and Proca field perturbations~\cite{BarraganAmado:2018zpa,BarraganAmado:2021uyw,Amado:2021erf,BarraganAmado:2025pfy}. Meanwhile, semiclassical conformal blocks were introduced to solve the connection problems for Heun equations~\cite{Bonelli:2021uvf,Bonelli:2022ten}. As a result, connection coefficients related to solutions around different singular points of the Heun equation were exploited to compute thermal correlators, greybody factors, quasinormal modes, and Love numbers in these five-dimensional backgrounds~\cite{Dodelson:2022yvn,He:2023wcs,Jia:2024zes,BarraganAmado:2024tfu,Arnaudo:2024sen,Ren:2024hwf}. In four dimensions, the so-called Heun-gravity correspondence has achieved extraordinary success in describing scattering processes~\cite{CarneirodaCunha:2015hzd,Bonelli:2021uvf} and revealing new features of the quasinormal mode spectrum~\cite{Cavalcante:2024swt,Cavalcante:2024kmy,Cavalcante:2025abr}.

On the other hand, ODEs with five singular points have been less investigated, and their underlying analytic aspects remain largely unexplored. In this regard, a remarkable step was made in~\cite{CarneirodaCunha:2021jsu}, where the theory of isomonodromic deformations was utilized to solve the accessory parameter problem associated with conformal maps to polycircular arc domains with any number of vertices. The solution was expressed in terms of the isomonodromic tau function introduced by Gavrylenko and Lisovyy~\cite{Gavrylenko:2016zlf}. As an explicit example, the authors considered the particular case of five vertices. Then, using the isomonodromic tau function for five regular singular points, an implementation for the computation of the quasinormal modes of scalar fields in accelerating Kerr--Newman--AdS black holes was presented in~\cite{BarraganAmado:2023wxt}. More recently, semiclassical $n$-point conformal blocks were introduced to study Fuchsian equations with five and six regular singular points~\cite{Liu:2024eut,Arnaudo:2025kof}.

In this work, we compute the quasinormal modes (QNMs) of a massive scalar field in a seven-dimensional rotating black hole in asymptotically anti-de Sitter spacetime, specifically the Kerr--AdS$_{7}$ black hole. The motivation for studying linear scalar perturbations of this spacetime is twofold. First, the separation of variables reduces the wave equation to a system of three second-order linear ODEs~\cite{Frolov:2006pe}. In particular, this background allows us to explore Fuchsian differential equations with five singularities, focusing on the relation between their accessory parameters and semiclassical conformal blocks. See Refs.~\cite{Lencses:2017dgf,Lisovyy:2021bkm,daCunha:2022ewy} for accessory parameter expansions in terms of semiclassical conformal blocks for the case of Heun equations. Second, in the framework of the AdS/CFT correspondence, the computation of the quasinormal frequencies correspond to the poles of the retarded thermal correlators of the associated dual operators on the conformal side~\cite{Nunez:2003eq}. Thus, perturbations of the Kerr--AdS$_{7}$ black hole could provide new insights into the six-dimensional $\mathcal{N}=(2,0)$ superconformal field theory dual to M-theory on AdS$_{7}$ $\times$ S$^{4}$. 

This paper is organized as follows. In Section~\ref{sec:2}, we introduce the metric of the Kerr--AdS$_{7}$ black hole with three independent rotation parameters. We then derive the angular and radial ODEs and recast them as Fuchsian differential equations on the Riemann sphere. In Section~\ref{sec:3}, we compute perturbative expansions for the accessory parameters using Hill determinants. These results are used in Section~\ref{sec:3.1} to obtain analytic expressions for the angular eigenvalues in the slowly rotating limit. In Section~\ref{sec:3.2}, we perform the computation of the quasinormal modes in the small horizon radius limit. Finally, we summarize our results and present our conclusions in Section~\ref{sec:4}.

Several technical details and complementary computations are presented in the appendices. Appendix~\ref{sec:A} contains higher order coefficients of the accessory parameter expansions and of the expansions of the separation constants. In Appendix~\ref{sec:B}, we compute the angular eigenvalues using accessory parameter expansions in the regime $0 < \vert t \vert < 1 < \vert q \vert < \infty$. Appendix~\ref{sec:C} presents an alternative derivation of the QNM frequencies based on a different M\"{o}bius transformation, corresponding to accessory parameter expansions in the regime $0 < \vert t \vert < \vert q \vert < 1$.

\section{Scalar perturbations in the Kerr--AdS\texorpdfstring{$_{7}$}{7} spacetime}
\label{sec:2}

\subsection{The metric of the seven-dimensional Kerr--AdS black hole}
\label{sec:2.1}
Our starting point is the general Kerr--NUT--AdS metric in odd dimensions $D = 2n+1$ obtained in~\cite{Chen:2006xh}. The solution is parametrized by a mass parameter $M$, $\lfloor\frac{D-1}{2}\rfloor$ rotation parameters $a_{i}$, and $\lfloor\frac{D-2}{2}\rfloor$ NUT parameters $L_{\alpha}$. The metric is then given by
\begin{subequations}
    \begin{align}\label{eq:d_dim_Kerr_NUT_AdS}
        ds^{2} = \frac{U}{X}dr^{2} &+ \sum_{\alpha=1}^{n-1}\frac{U_{\alpha}}{X_{\alpha}}dy_{\alpha}^{2} - \frac{X}{U}\left[\frac{W}{\prod_{i}^{n}\Xi_{i}} dt - \sum_{i=1}^{n}a_{i}^{2}\gamma_{i}\frac{d\phi_{i}}{\epsilon_{i}}\right]^{2}\nonumber\\ 
        &+ \sum_{\alpha=1}^{n-1}\frac{X_{\alpha}}{U_{\alpha}}\left[\frac{\bigl(1+\tfrac{r^{2}}{L^{2}}\bigr)W}{\left(1-\tfrac{y_{\alpha}^{2}}{L^{2}}\right)\prod_{i}^{n}\Xi_{i}}dt - \sum_{i=1}^{n}\frac{a_{i}^{2}(r^{2} + a_{i}^{2})\gamma_{i}}{a_{i}^{2} -  y_{\alpha}^{2}}\frac{d\phi_{i}}{\epsilon_{i}}\right]^{2}\\
        &\qquad\qquad+\frac{\prod_{k=1}^{n}a_{k}^{2}}{r^{2}\prod_{\alpha=1}^{n-1}y_{\alpha}^{2}}\left[\frac{\bigl(1 + \tfrac{r^{2}}{L^{2}} \bigr)W}{\prod_{i}^{n}\Xi_{i}}dt - \sum_{i=1}^{n}\left(r^{2} + a_{i}^{2}\right)\gamma_{i}\frac{d\phi_{i}}{\epsilon_{i}}\right]^{2}\,,\nonumber
    \end{align}
where
    \begin{align}\label{eq:d_dim_functions}
        &U = \prod_{\alpha=1}^{n-1}\left(r^{2} + y_{\alpha}^{2}\right)\,, \qquad U_{\alpha} = - \left(r^{2} + y_{\alpha}^{2}\right)\prod_{\beta=1}^{\prime\,n-1}\left(y_{\beta}^{2}-y_{\alpha}^{2}\right)\,, \quad 1 \leq \alpha \leq n-1\,,\nonumber\\
        &W = \prod_{\alpha=1}^{n-1}\biggl(1 - \frac{y_{\alpha}^{2}}{L^{2}}\biggr)\,, \qquad \Xi_{i} = 1 - \frac{a_{i}^{2}}{L^{2}}\,,\nonumber\\
        &\epsilon_{i} = a_{i}\Xi_{i}\prod_{k=1}^{'n}\left(a_{i}^{2} - a_{k}^{2}\right)\,, \qquad \gamma_{i} = \prod_{\alpha=1}^{n-1}\left(a_{i}^{2} - y_{\alpha}^{2}\right)\,, \qquad 1 \leq i \leq n\,,\\
        &X = \frac{1 + \tfrac{r^{2}}{L^{2}}}{r^{2}}\prod_{k=1}^{n}\left(r^{2} + a_{k}^{2}\right)  - 2M\,,\nonumber\\
        &X_{\alpha} = \frac{1 - \tfrac{y_{\alpha}^{2}}{L^{2}}}{y_{\alpha}^{2}}\prod_{k=1}^{n}\left(a_{k}^{2} - y_{\alpha}^{2}\right)+2L_{\alpha}\,, \qquad 1\leq\alpha\leq n-1\,.\nonumber
    \end{align}
\end{subequations}
Note that the notation $\prod^{'}$ indicates that the term in the full product that vanishes is to be omitted. Without loss of generality, we may order the rotation parameters such that $a_{1} \leq a_{2} \leq \cdots \leq a_{n}$, in which case the compact coordinates $y_{\alpha}$ must lie in the ranges $a_{\alpha} \leq y_{\alpha} \leq a_{\alpha+1}$. Furthermore, the time and azimuthal coordinates can be rescaled as follows
\begin{equation}\label{eq:tilde_coord}
    t = \tilde{t}\prod_{i=1}^{n}\Xi_{i}\,, \qquad \phi_{i} = \epsilon_{i}\tilde{\phi}_{i}\,,
\end{equation}
which simplifies the metric but changes the normalization of $\tilde{t}$ and modifies the period of each $\tilde{\phi}_{i}$ coordinate, since the coordinates $\phi_{i}$ have period $2\pi$. 

As we are primarily concerned with Kerr--AdS metrics, the NUT parameters $L_{\alpha}$ are set to zero, \textit{i.e.} $L_{\alpha} = 0$ for all $1 \leq \alpha \leq n-1$. For the specific case $n=3$, the seven-dimensional Kerr--AdS black hole solution in terms of the coordinates $(\tilde{t},r,y_{\alpha},\tilde{\phi}_{i})$ is given by
\begin{subequations}
    \begin{equation}\label{eq:metric}
        \begin{split}
                ds^{2} &= \frac{U}{X}dr^{2} + \frac{U_{1}}{X_{1}}dy_{1} ^{2} + \frac{U_{2}}{X_{2}}dy_{2}^{2} - \frac{X}{U}\left[W d\tilde{t} - a_{1}^{2}\gamma_{1}d\tilde{\phi}_{1} - a_{2}^{2}\gamma_{2}d\tilde{\phi}_{2} - a_{3}^{2}\gamma_{3}d\tilde{\phi}_{3}\right]^{2}\\ 
                &+ \frac{X_{1}}{U_{1}}\Biggl[\frac{\bigl(1+\tfrac{r^{2}}{L^{2}}\bigr)W}{1-\tfrac{y_{1}^{2}}{L^{2}}}d\tilde{t} - \frac{a_{1}^{2}(r^{2} + a_{1}^{2})\gamma_{1}}{a_{1}^{2} - y_{1}^{2}}d\tilde{\phi}_{1} - \frac{a_{2}^{2}(r^{2} + a_{2}^{2})\gamma_{2}}{a_{2}^{2} - y_{1}^{2}}d\tilde{\phi}_{2} - \frac{a_{3}^{2}(r^{2} + a_{3}^{2})\gamma_{3}}{a_{3}^{2} - y_{1}^{2}}d\tilde{\phi}_{3}\Biggr]^{2}\\
                &+ \frac{X_{2}}{U_{2}}\Biggl[\frac{\bigl(1+\tfrac{r^{2}}{L^{2}}\bigr)W}{1-\tfrac{y_{2}^{2}}{L^{2}}}d\tilde{t} - \frac{a_{1}^{2}(r^{2} + a_{1}^{2})\gamma_{1}}{a_{1}^{2} - y_{2}^{2}}d\tilde{\phi}_{1} - \frac{a_{2}^{2}(r^{2} + a_{2}^{2})\gamma_{2}}{a_{2}^{2} - y_{2}^{2}}d\tilde{\phi}_{2} - \frac{a_{3}^{2}(r^{2} + a_{3}^{2})\gamma_{3}}{a_{3}^{2} - y_{2}^{2}}d\tilde{\phi}_{3}\Biggr]^{2}\\
                &+ \frac{a_{1}^{2}a_{2}^{2}a_{3}^{2}}{r^{2}y_{1}^{2}y_{2}^{2}}\left[\bigl(1 + \tfrac{r^{2}}{L^{2}}\bigr)Wd\tilde{t} -(r^{2} + a_{1}^{2})\gamma_{1}d\tilde{\phi}_{1} - (r^{2} + a_{2}^{2})\gamma_{2}d\tilde{\phi}_{2} - (r^{2} + a_{3}^{2})\gamma_{3}d\tilde{\phi}_{3}\right]^{2}\,,
        \end{split}
    \end{equation}
where the functions $U,U_{1},U_{2},W$ and $\gamma_{i}\,(i=1,2,3)$ are defined in \eqref{eq:d_dim_functions}. A more convenient way to write $X$, $X_{1}$ and $X_{2}$ is 
    \begin{equation}\label{eq:X_functions}
        \begin{split}
            X &= \frac{1}{r^{2}}\biggl(1 + \frac{r^{2}}{L^{2}}\biggr)\left(r^{2} + a_{1}^{2}\right)\left(r^{2} + a_{2}^{2}\right)\left(r^{2} + a_{3}^{2}\right) - 2M\\ 
            &=  \frac{r^{6}} {L^{2}} + C_{0}r^{4} + C_{1}r^{2} + C_{2} - 2M + \frac{C_{3}}{r^{2}}\,,\\
            X_{1} &= \frac{1}{y_{1}^{2}}\biggl(1 - \frac{y_{1}^{2}}{L^{2}}\biggr)\left(a_{1}^{2} - y_{1}^{2}\right)\left(a_{2}^{2} - y_{1}^{2}\right)\left(a_{3}^{2} - y_{1}^{2}\right)\\
            &= \frac{y_{1}^{6}}{L^{2}} - C_{0}y_{1}^{4} + C_{1}y_{1}^{2} - C_{2} + \frac{C_{3}}{y_{1}^{2}}\,,\\
            X_{2} &= \frac{1}{y_{2}^{2}}\biggl(1 - \frac{y_{2}^{2}}{L^{2}}\biggr)\left(a_{1}^{2} - y_{2}^{2}\right)\left(a_{2}^{2} - y_{2}^{2}\right)\left(a_{3}^{2} - y_{2}^{2}\right)\\
            &= \frac{y_{2}^{6}}{L^{2}} - C_{0}y_{2}^{4} + C_{1}y_{2}^{2} - C_{2} + \frac{C_{3}}{y_{2}^{2}}\,.
        \end{split}
    \end{equation}
\end{subequations}
Here the $C_{k}\,(k=0,1,2,3)$ are constants, expressed in terms of the rotation parameters $a_{i}\,(i=1,2,3)$ and the AdS radius $L$, given by
\begin{equation}
    \begin{split}
        &C_{0} = 1 + \frac{a_{1}^{2}}{L^{2}} + \frac{a_{2}^{2}}{L^{2}} + \frac{a_{3}^{2}}{L^{2}}\,,\\
        &C_{1} = a_{1}^{2}\biggl(1+\frac{a_{2}^{2}}{L^{2}}\biggr) + a_{2}^{2}\biggl(1 + \frac{a_{3}^{2}}{L^{2}}\biggr) + a_{3}^{2}\biggl(1 + \frac{a_{1}^{2}}{L^{2}}\biggr)\,,\\
        &C_{2} = a_{1}^{2}a_{2}^{2} + a_{1}^{2}a_{3}^{2} + a_{2}^{2}a_{3}^{2} + \frac{a_{1}^{2}a_{2}^{2}a_{3}^{2}}{L^{2}}\,, \\
        &C_{3} = a_{1}^{2}a_{2}^{2}a_{3}^{2}\,.
    \end{split}
\end{equation}
Furthermore, the metric in Eq.~\eqref{eq:metric} can be rewritten as
\begin{equation}\label{eq:new_metric}
    \begin{split}
        ds^{2} &= \frac{U}{X}dr^{2} + \frac{U_{1}}{X_{1}}dy_{1} ^{2} + \frac{U_{2}}{X_{2}}dy_{2}^{2} - \frac{X}{U}\left[d\psi_{0} - \bigl(y_{1}^{2} + y_{2}^{2}\bigr)d\psi_{1} + y_{1}^{2}y_{2}^{2}d\psi_{2}\right]^{2}\\ 
        &+ \frac{X_{1}}{U_{1}}\left[d\psi_{0} + \bigl(r^{2} -y_{2}^{2}\bigr)d\psi_{1} - r^{2}y_{2}^{2}d\psi_{2}\right]^{2} + \frac{X_{2}}{U_{2}}\left[d\psi_{0} + \bigl(r^{2} - y_{1}^{2} \bigr)d\psi_{1} - r^{2}y_{1}^{2}d\psi_{2}\right]^{2}\\
        &+ \frac{C_{3}}{r^{2}y_{1}^{2}y_{2}^{2}}\left[d\psi_{0} + \bigl(r^{2} - y_{1}^{2} - y_{2}^{2}\bigr)d\psi_{1} + \bigl(y_{1}^{2}y_{2}^{2} - r^{2}y_{1}^{2} - r^{2}y_{2}^{2}\bigr)d\psi_{2} + r^{2}y_{1}^{2}y_{2}^{2}d\psi_{3}\right]^{2}\,,
    \end{split}
\end{equation}
where $\psi_{k}\,(k=0,1,2,3)$ are related to the original asymptotically static coordinates $t$ and $\phi_{i}\,(i=1,2,3)$ defined in \eqref{eq:tilde_coord} by
\begin{equation}\label{eq:time_and_phis}
    \begin{split}
        t &= \psi_{0} - \left(a_{1}^{2} + a_{2}^{2} + a_{3}^{2}\right)\psi_{1} + \left(a_{1}^{2}a_{2}^{2} + a_{1}^{2}a_{3}^{2} + a_{2}^{2}a_{3}^{2}\right)\psi_{2} - a_{1}^{2}a_{2}^{2}a_{3}^{2}\psi_{3}\,,\\
        \frac{\phi_{1}}{a_{1}} &= \frac{\psi_{0}}{L^{2}} - \biggl(1 + \frac{a_{2}^{2}}{L^{2}} + \frac{a_{3}^{2}}{L^{2}}\biggr)\psi_{1} + \biggl(a_{2}^{2} + a_{3}^{2} + \frac{a_{2}^{2}a_{3}^{2}}{L^{2}} \biggr)\psi_{2} - a_{2}^{2}a_{3}^{2}\psi_{3}\,,\\
        \frac{\phi_{2}}{a_{2}} &= \frac{\psi_{0}}{L^{2}} - \biggl(1 + \frac{a_{1}^{2}}{L^{2}} + \frac{a_{3}^{2}}{L^{2}}\biggr)\psi_{1} + \biggl(a_{1}^{2} + a_{3}^{2} + \frac{a_{1}^{2}a_{3}^{2}}{L^{2}}\biggr)\psi_{2} - a_{1}^{2}a_{3}^{2}\psi_{3}\,,\\
        \frac{\phi_{3}}{a_{3}} &= \frac{\psi_{0}}{L^{2}} - \biggl(1 + \frac{a_{1}^{2}}{L^{2}} + \frac{a_{2}^{2}}{L^{2}}\biggr)\psi_{1} + \biggl(a_{1}^{2} + a_{2}^{2} + \frac{a_{1}^{2}a_{2}^{2}}{L^{2}} \biggr)\psi_{2} - a_{1}^{2}a_{2}^{2}\psi_{3}\,,
    \end{split}
\end{equation}
and the determinant of the metric gives 
\begin{equation}
    \begin{split}
        g &= \det g_{\mu\nu}\\
        &= -C_{3}r^{2}y_{1}^{2}y_{2}^{2}\bigl(r^{2} + y_{1}^{2}\bigr)^{2}\bigl(r^{2} + y_{2}^{2}\bigr)^{2}\bigl(y_{1}^{2} - y_{2}^{2}\bigr)^{2}\,.
    \end{split}
\end{equation}

\subsection{Separability of the scalar field equation}
\label{sec:2.2}
The Klein--Gordon (KG) equation for a scalar field of mass $m$ is governed by
\begin{equation}\label{eq:scalar_wave}
    \frac{1}{\sqrt{-g}}\partial_{\mu}\bigl(\sqrt{-g}g^{\mu\nu}\partial_{\nu}\bigr)\Phi - m^{2}\Phi = 0\,.
\end{equation}
The separability of the scalar field equation in the general Kerr--NUT--AdS spacetimes was proved in \cite{Frolov:2006pe}. Therefore, the KG equation~\eqref{eq:scalar_wave} in the Kerr-AdS$_{7}$ background~\eqref{eq:new_metric} admits a multiplicative separation of variables
\begin{equation}
    \Phi = R(r)Y_{1}(y_{1})Y_{2}(y_{2})\prod_{k=0}^{3}e^{i\Psi_{k}\psi_{k}}\,,
\end{equation}
where $R(r)$ is the radial function and $Y_{i}(y_{i})$ are the angular functions. Substituting this ansantz into Eq.~\eqref{eq:scalar_wave} yields 
\begin{equation}\label{eq:KG}
    \begin{aligned}
        &\bigl(y_{1}^{2} - y_{2}^{2}\bigr)\left[\frac{1}{r R(r)}\partial_{r}\left(r X(r)\partial_{r}R(r)\right) + \frac{1}{X(r)}\left(r^{4}\Psi_{0} - r^{2}\Psi_{1} + \Psi_{2} - \frac{\Psi_{3}}{r^{2}}\right)^{2} - \frac{1}{C_{3}}\frac{\Psi_{3}^{2}}{r^{2}}\right]\\
        &+ \bigl(r^{2}+y_{2}^{2}\bigr)\left[\frac{1}{y_{1}Y_{1}(y_{1})}\partial_{y_{1}}\left(y_{1}X_{1}(y_{1})\partial_{y_{1}}Y_{1}(y_{1})\right) - \frac{1}{X_{1}(y_{1})}\left(y_{1}^{4}\Psi_{0} + y_{1}^{2}\Psi_{1} + \Psi_{2} + \frac{\Psi_{3}}{y_{1}^{2}}\right)^{2} + \frac{1}{C_{3}}\frac{\Psi_{3}^{2}}{y_{1}^{2}}\right]\\ 
        &- \bigl(r^{2}+y_{1}^{2}\bigr)\Biggl[\frac{1}{y_{2}Y_{2}(y_{2})}\partial_{y_{2}}\left(y_{2}X_{2}(y_{2})\partial_{y_{2}}Y_{2}(y_{2})\right) - \frac{1}{X_{2}(y_{2})}\left(y_{2}^{4}\Psi_{0} + y_{2}^{2}\Psi_{1} + \Psi_{2} + \frac{\Psi_{3}}{y_{2}^{2}}\right)^{2}
        + \frac{1}{C_{3}}\frac{\Psi_{3}^{2}}{y_{2}^{2}}\Biggr]\\
        &- m^{2}\bigl(r^{2}+y_{1}^{2}\bigr)\bigl(r^{2}+y_{2}^{2}\bigr)\bigl(y_{1}^{2} - y_{2}^{2}\bigr)=0\,.
    \end{aligned}
\end{equation}
The constants $\Psi_{k}$ follow from requiring $e^{i\sum_{k=0}^{3}\Psi_{k}\psi_{k}} = e^{-i w t + i\sum_{i=1}^{3}m_{i}\phi_{i}}$, together with the explicit expressions for $\psi_{k}$ in terms of $(t,\phi_{i})$ from Eq.~\eqref{eq:time_and_phis}.
We then have 
\begin{equation}\label{eq:psis}
	\begin{split}
    	\Psi_{0} &= -w + \frac{m_{1}a_{1}}{L^{2}} + \frac{m_{2}a_{2}}{L^{2}} + \frac{m_{3}a_{3}}{L^{2}}\,, \\
		\Psi_{1} &= \left(a_{1}^{2} + a_{2}^{2} + a_{3}^{2}\right)w - a_{1}\left(1 + \frac{a_{2}^{2}}{L^{2}} + \frac{a_{3}^{2}}{L^{2}}\right)m_{1} - a_{2}\left(1 + \frac{a_{1}^{2}}{L^{2}} + \frac{a_{3}^{2}}{L^{2}}\right)m_{2}\\
		&\qquad - a_{3}\left(1 + \frac{a_{1}^{2}}{L^{2}} + \frac{a_{2}^{2}}{L^{2}}\right)m_{3}\,, \\
		\Psi_{2} &= -\left(a_{1}^{2}a_{2}^{2} + a_{1}^{2}a_{3}^{2} + a_{2}^{2}a_{3}^{2}\right)w + a_{1}\left[a_{3}^{2} + a_{2}^{2}\left(1 + \frac{a_{3}^{2}}{L^{2}}\right)\right]m_{1}\\ 
		&\qquad + a_{2}\left[a_{3}^{2} + a_{1}^{2}\left(1 + \frac{a_{3}^{2}}{L^{2}}\right)\right]m_{2} + a_{3}\left[a_{2}^{2} + a_{1}^{2}\left(1 + \frac{a_{2}^{2}}{L^{2}}\right)\right]m_{3}\,, \\ 
		\Psi_{3} &= a_{1}^{2}a_{2}^{2}a_{3}^{2}\left(w - \frac{m_{1}}{a_{1}} - \frac{m_{2}}{a_{2}} - \frac{m_{3}}{a_{3}}\right)\,. 
	\end{split}
\end{equation}
where $w \in \mathbb{C}$ is the mode frequency and $m_{i}\,(i=1,2,3) \in \mathbb{Z}$ are the azimuthal numbers. Furthermore, Eq.~\eqref{eq:KG} can be separated into three ordinary second--order differential equations 
\begin{subequations}\label{eq:three_odes}
    \begin{equation}\label{eq:rad}
        \begin{split}
            \frac{1}{r R(r)}\frac{d}{d r}\left(r X(r)\frac{d R}{dr}\right) + \frac{1}{X(r)}\biggl(r^{4}\Psi_{0} &- r^{2}\Psi_{1} + \Psi_{2} - \frac{\Psi_{3}}{r^{2}}\biggr)^{2}\\ 
            &- \frac{1}{C_{3}}\frac{\Psi_{3}^{2}}{r^{2}} - m^{2}r^{4} - b_{1}r^{2} - b_{2}=0\,,
        \end{split}
    \end{equation}
    \begin{equation}\label{eq:ang1}
        \begin{split}
            \frac{1}{y_{1}Y_{1}(y_{1})}\frac{d}{dy_{1}}\left(y_{1}X_{1}(y_{1})\frac{d Y_{1}}{d y_{1}}\right) - \frac{1}{X_{1}(y_{1})}\biggl(y_{1}^{4}\Psi_{0} &+ y_{1}^{2}\Psi_{1} + \Psi_{2} + \frac{\Psi_{3}}{y_{1}^{2}}\biggr)^{2}\\ 
            &+ \frac{1}{C_{3}}\frac{\Psi_{3}^{2}}{y_{1}^{2}} - m^{2}y_{1}^{4} + b_{1}y_{1}^{2} - b_{2}=0\,,
        \end{split}
    \end{equation}
    \begin{equation}\label{eq:ang2}
        \begin{split}
            \frac{1}{y_{2}Y_{2}(y_{2})}\frac{d}{dy_{2}}\left(y_{2}X_{2}(y_{2})\frac{d Y_{2}}{d y_{2}}\right) - \frac{1}{X_{2}(y_{2})}\biggl(y_{2}^{4}\Psi_{0} &+ y_{2}^{2}\Psi_{1} + \Psi_{2} + \frac{\Psi_{3}}{y_{2}^{2}}\biggr)^{2}\\ 
            &+ \frac{1}{C_{3}}\frac{\Psi_{3}^{2}}{y_{2}^{2}} - m^{2}y_{2}^{4} + b_{1}y_{2}^{2} - b_{2} = 0\,,
        \end{split}
    \end{equation}
\end{subequations}
where $b_{1}$ and $b_{2}$ are the separation constants and correspond to the angular eigenvalues. In the next subsection, we apply a series of transformations to the radial and angular equations \eqref{eq:three_odes} to express them as Fuchsian equations on the Riemann sphere. 

\subsubsection{Radial system}
\label{sec:2.2.1}
The equation for $R(r)$ reads
\begin{subequations}
\begin{equation}\label{eq:radial}
    \begin{split}
        \frac{1}{r}\frac{d}{dr}\left(r X(r)\frac{d R}{dr}\right) + \biggl[ \frac{1}{X(r)}\biggl(r^{4}\Psi_{0} &- r^{2}\Psi_{1} + \Psi_{2} - \frac{1}{r^{2}}\Psi_{3}\biggr)^{2}\\
        &\quad- \frac{1}{C_{3}}\frac{\Psi_{3}^{2}}{r^{2}} - m^{2}r^{4} - b_{1}r^{2} - b_{2}\biggr]R(r) = 0\,,
    \end{split}
\end{equation}
where the metric function $X(r)$ given in \eqref{eq:X_functions} can be rewritten as
\begin{equation}
    X(r) = \frac{\left(r^{2} - r_{3}^{2}\right)\left(r^{2} - r_{2}^{2}\right)\left(r^{2} - r_{1}^{2}\right)\left(r^{2} - r_{0}^{2}\right)}{L^{2}r^{2}}\,.
\end{equation}
\end{subequations}
For convenience, we introduce a dimensionless radial coordinate
\begin{equation}\label{eq:dim_coordinate}
    u = \frac{r}{L}\,, \qquad u_{k} = \frac{r_{k}}{L}\,, \qquad k = 0,1,2,3\,,
\end{equation}
such that $X(r) = L^{4}f(u) = L^{4}\frac{\left(u^{2} - u_{3}^{2}\right)\left(u^{2} - u_{2}^{2}\right)\left(u^{2} - u_{1}^{2}\right)\left(u^{2} - u_{0}^{2}\right)}{u^{2}}$. Then, by rescaling the physical parameters and the separation constants as follows
\begin{equation}\label{eq:dim_vars_params}
    \omega = Lw\,, \quad \alpha_{1} = \frac{a_{1}}{L}\, \quad \alpha_{2} = \frac{a_{2}}{L}\,, \quad \alpha_{3} = \frac{a_{3}}{L}\,, \quad \mu = L m\,, \quad \beta_{1} = b_{1}\,, \quad \beta_{2} = \frac{b_{2}}{L^{2}}\,, 
\end{equation}
one obtains that $\Psi_{k}$ and $C_{3}$ scale 
\begin{subequations}\label{eq:dim_psis}
\begin{equation}
    \begin{split}
        \Psi_{0} = \frac{\varPsi_{0}}{L}\,, \quad \Psi_{1} = L\varPsi_{1}\,, \quad
        \Psi_{2} = L^{3}\varPsi_{2}\,, \quad \Psi_{3} = L^{5}\varPsi_{3}\,, \quad
        C_{3} = L^{6}\chi_{3}\,,
    \end{split}
\end{equation}
where
\begin{equation}\label{eq:varPsis}
    \begin{split}
        &\varPsi_{0} = -\omega + m_{1}\alpha_{1} + m_{2}\alpha_{2} + m_{3}\alpha_{3}\,,\\
        &\varPsi_{1} = \left(\alpha_{1}^{2} + \alpha_{2}^{2} + \alpha_{3}^{2}\right)\omega - m_{1}\alpha_{1}\left(1 + \alpha_{2}^{2} + \alpha_{3}^{2}\right) \\ 
        &\qquad\qquad - m_{2}\alpha_{2}\left(1 + \alpha_{1}^{2} + \alpha_{3}^{2}\right) - m_{3}\alpha_{3}\left(1 + \alpha_{1}^{2} + \alpha_{2}^{2}\right)\,,\\
        &\varPsi_{2} = -\left(\alpha_{1}^{2}\alpha_{2}^{2} + \alpha_{1}^{2}\alpha_{3}^{2} + \alpha_{2}^{2}\alpha_{3}^{2}\right)\omega + m_{1}\alpha_{1}\left(\alpha_{2}^{2} + \alpha_{3}^{2} + \alpha_{2}^{2}\alpha_{3}^{2}\right)\\
        &\qquad\qquad + m_{2}\alpha_{2}\left(\alpha_{1}^{2} + \alpha_{3}^{2} + \alpha_{1}^{2}\alpha_{3}^{2}\right) + m_{3}\alpha_{3}\left(\alpha_{1}^{2} + \alpha_{2}^{2} + \alpha_{1}^{2}\alpha_{2}^{2}\right)\,,\\
        &\varPsi_{3} = \alpha_{1}^{2}\alpha_{2}^{2}\alpha_{3}^{2}\left(\omega - \frac{m_{1}}{\alpha_{1}} - \frac{m_{2}}{\alpha_{2}} - \frac{m_{3}}{\alpha_{3}} \right)\,,\\
        &\chi_{3} = \alpha_{1}^{2}\alpha_{2}^{2}\alpha_{3}^{2}\,,
    \end{split}
\end{equation}
\end{subequations}
yielding an ODE that does not depend on the AdS radius $L$. In terms of the new variable, the radial equation takes the following form
\begin{equation}\label{eq:dim_radial}
    \begin{split}
        \frac{d^{2}R}{du^{2}} + \left(\frac{f^{\prime}(u)}{f(u)} + \frac{1}{u}\right)\frac{d R}{du} + \Biggl[&\frac{\left( u^{6}\varPsi_{0} - u^{4}\varPsi_{1} + u^{2}\varPsi_{2} - \varPsi_{3}\right)^{2}}{u^{4} f(u)^{2}} \\
        &\qquad - \frac{1}{f(u)}\left(u^{4}\mu^{2} + u^{2}\beta_{1} + \beta_{2} + \frac{\varPsi_{3}^{2}}{u^{2}\chi_{3}}\right)\Biggr]R(u) = 0\,,
    \end{split}
\end{equation}
which can be rewritten as
\begin{subequations}
\begin{equation}\label{eq:dim_radial_ode}
    \begin{gathered}
        \frac{d^{2}R}{du^{2}} + \left(\frac{f^{\prime}(u)}{f(u)} + \frac{1}{u}\right)\frac{d R}{du} + \Biggl[\frac{2u_{0}A_{0}}{u^{2} - u_{0}^{2}} + \frac{2u_{1}A_{1}}{u^{2} - u_{1}^{2}} + \frac{2u_{2}A_{2}}{u^{2} - u_{2}^{2}} +\frac{2u_{3}A_{3}}{u^{2} - u_{3}^{2}}\\ 
        +\frac{2B_{0}\left(u^{2} + u_{0}^{2}\right)}{\left(u^{2} - u_{0}^{2}\right)^{2}} + \frac{2B_{1}\left(u^{2} + u_{1}^{2}\right)}{\left(u^{2} - u_{1}^{2}\right)^{2}} +\frac{2B_{2}\left(u^{2} + u_{2}^{2}\right)}{\left(u^{2} - u_{2}^{2}\right)^{2}} +\frac{2B_{3}\left(u^{2} + u_{3}^{2}\right)}{\left(u^{2} - u_{3}^{2}\right)^{2}}\\
        \qquad - \frac{1}{f(u)}\left(u^{4}\mu^{2} + u^{2}\beta_{1} + \beta_{2} + \frac{\varPsi_{3}^{2}}{u^{2}\chi_{3}}\right)\Biggr]R(u) = 0\,,
    \end{gathered}
\end{equation}
where
\begin{equation}\label{eq:coeffs_rad}
    \begin{split}
        &A_{k} = \lim_{u \to u_{k}}\frac{d}{d u}\left[\frac{\left(u - u_{k} \right)^{2}\left(u^{6}\varPsi_{0} - u^{4}\varPsi_{1} + u^{2}\varPsi_{2} - \varPsi_{3}\right)^{2}}{\left(u^{2} - u_{0}^{2}\right)^{2}\left(u^{2} - u_{1}^{2}\right)^{2}\left(u^{2} - u_{2}^{2}\right)^{2}\left(u^{2} - u_{3}^{2}\right)^{2}}\right]\,,\\
        &B_{k} = \lim_{u \to u_{k}}\frac{\left(u - u_{k} \right)^{2}\left(u^{6}\varPsi_{0} - u^{4}\varPsi_{1} + u^{2}\varPsi_{2} - \varPsi_{3}\right)^{2}}{\left(u^{2} - u_{0}^{2}\right)^{2}\left(u^{2} - u_{1}^{2}\right)^{2}\left(u^{2} - u_{2}^{2}\right)^{2}\left(u^{2} - u_{3}^{2}\right)^{2}}\,, \qquad k = 0,1,2,3\,.
    \end{split}
\end{equation}
\end{subequations}
The radial equation~\eqref{eq:dim_radial_ode} possesses five regular singular points in the $u^{2}$ variable, located at $u_{0}^{2}, u_{1}^{2},u_{2}^{2},u_{3}^{2}, \infty$. One can recast this equation via the following M\"{o}bius transformation
\begin{equation}\label{eq:Mobius}
    z = \frac{u_{1}^{2} - u_{0}^{2}}{u_{1}^{2} - u_{2}^{2}}\frac{u^{2} - u_{2}^{2}}{u^{2} - u_{0}^{2}}\,,
\end{equation}
which maps the singular points on the Riemann sphere as follows
\begin{subequations}
\begin{equation}\label{eq:z_to_u}
    \left( u_{0}^{2}, u_{1}^{2}, u_{2}^{2}, u_{3}^{2}, \infty \right) \,\longmapsto \, \left( \infty, 1, 0, t, q\right)\,,
\end{equation}
with
\begin{equation}\label{eq:moduli_rad}
    t = \frac{u_{1}^{2} - u_{0}^{2}}{u_{1}^{2} - u_{2}^{2}}\frac{u_{3}^{2} - u_{2}^{2}}{u_{3}^{2} - u_{0}^{2}}, \qquad q = \frac{u_{1}^{2} - u_{0}^{2}}{u_{1}^{2} - u_{2}^{2}}\,.
\end{equation}
\end{subequations}
Near each singularity, one can expand the solution up to leading order as $R = a(z-z_{k})^{\rho_{z_k}^{-}} + b(z-z_{k})^{\rho_{z_k}^{+}}$, where $\rho_{z_k}^-$ and $\rho_{z_k}^+$ are the characteristic exponents at each singular point, with $a$ and $b$ being constants. The characteristic exponents of the Frobenius solutions near to each singularity are
\begin{subequations}
    \begin{equation}\label{eq:indicial_scalar}
        \begin{gathered}
            \rho^\pm_{0} = \pm i \sqrt{B_{2}} = \pm \frac{\theta_{u_{2}}}{2}\,,\qquad \rho^{\pm}_{t} = \pm i \sqrt{B_{3}} = \pm \frac{\theta_{u_{3}}}{2}\,, \qquad \rho^{\pm}_{1} = \pm i \sqrt{B_{1}} = \pm \frac{\theta_{u_{1}}}{2}\,,\\
            \rho^{\pm}_{q} = \frac{1}{2}\left(3 \pm \sqrt{9 + \mu^{2}}\right)\,, \qquad\rho^{\pm}_{\infty} = \pm i \sqrt{B_{0}} = \pm \frac{\theta_{u_{0}}}{2}\,,
        \end{gathered}
    \end{equation}
where 
    \begin{equation}
        \theta_{u_{k}} = i\frac{\left(u_{k}^{6}\varPsi_{0} - u_{k}^{4}\varPsi_{1} + u_{k}^{2}\varPsi_{2} - \varPsi_{3}\right)}{\sqrt{u_{k}^{2}\prod_{j=0, j \neq k}^{3}(u_{k}^{2} - u_{j}^{2})^{2}}}\,, \qquad k =  0,1,2,3\,,
    \end{equation}
which in terms of the physical parameters, the characteristic exponents read
    \begin{equation}\label{eq:thetas}
        \begin{split}
            \theta_{u_{k}} = &i\frac{(u_{k}^{2} + \alpha_{1}^{2})(u_{k}^{2} + \alpha_{2}^{2})(u_{k}^{2} + \alpha_{3}^{2})}{\sqrt{u_{k}^{2}\prod_{j=0, j \neq k}^{3}(u_{k}^{2} - u_{j}^{2})^{2}}}\\
            &\qquad\qquad\times\left(\omega - \frac{m_{1}\alpha_{1}(1+u_{k}^{2})}{u_{k}^{2} + \alpha_{1}^{2}} - \frac{m_{2}\alpha_{2}(1+u_{k}^{2})}{u_{k}^{2} + \alpha_{2}^{2}} - \frac{m_{3}\alpha_{3}(1+u_{k}^{2})}{u_{k}^{2} + \alpha_{3}^{2}}\right)\,.
        \end{split}
    \end{equation}
\end{subequations}
Now, let us introduce the following $s$-homotopic transformation
\begin{equation}
    R(z) = z^{-1/2}(z-t)^{-1/2}(z-1)^{-1/2}(z-q)\psi(z)\,,
\end{equation}
such that the differential equation satisfied by $\psi(z)$ is 

\begin{equation}\label{eq:fuchsian_rad}
	\begin{split}
		\frac{d^{2}\mathcal{\psi}}{dz^{2}} + \biggl[\frac{\delta_{0}}{z^{2}} &+ \frac{\delta_{t}}{(z - t)^{2}} +  \frac{\delta_{1}}{(z - 1)^{2}} +\frac{\delta_{q}}{(z - q)^{2}} +                 \frac{\delta_{\infty} - \delta_{0} - \delta_{t} - \delta_{1} - \delta_{q}}{z(z - 1)}\\
    	&\qquad\qquad + \frac{(t-1)E_{t}}{z(z - t)(z - 1)} + \frac{(q-1)E_{q}}{z(z - q)(z - 1)}\biggr]\psi(z) = 0,
	\end{split}
\end{equation}
where
\begin{equation}\label{eq:delta_radial}
    \begin{gathered}
	        \delta_{0} = \frac{1}{4}\left(1 - \theta_{u_{2}}^{2}\right)\,, \quad \delta_{t} = \frac{1}{4}\left(1 - \theta_{u_{3}}^{2}\right)\,, \quad \delta_{1} = \frac{1}{4}\left(1 - \theta_{u_{1}}^{2}\right)\,,\\ 
            \delta_{q} = \frac{1}{4}\left(1 - (9 + \mu^{2}) \right)\,,
            \quad \delta_{\infty} = \frac{1}{4}\left(1 - \theta_{u_{0}}^{2}\right)\,,
    \end{gathered}
\end{equation}
\begin{subequations}\label{eq:accessory_radial}
    \begin{align}
        E_{t} &= -\frac{\left(u_{3}^{2}\beta_{1} + \beta_{2} + u_{3}^{4}\mu^{2} + \frac{\varPsi_{3}^{2}}{u_{3}^{2}\chi_{3}} \right)}{4(u_{0}^{2} - u_{2}^{2})(u_{1}^{2} - u_{3}^{2})} - \frac{\left(u_{3}^{2} - u_{2}^{2}\right)\left(u_{3}^{2} - u_{0}^{2}\right)}{2 u_{3}(u_{0}^{2} - u_{2}^{2})}A_{3} \label{eq:Et_radial}\\ 
        &\quad - \frac{\left(u_{3}^{2} - u_{2}^{2}\right)\left(u_{0}^{4} + 2u_{0}^{2}u_{3}^{2} - 3u_{3}^{4}\right)}{8 u_{3}^{2}\left(u_{0}^{2} - u_{2}^{2}\right)\left(u_{3}^{2} - u_{0}^{2}\right)}\theta_{u_{3}}^{2} + \frac{(u_{3}^{2} - u_{0}^{2})(u_{1}^{2} + u_{2}^{2} - 2u_{3}^{2})}{2(u_{0}^{2} - u_{2}^{2})(u_{1}^{2} - u_{3}^{2})} \,,\notag\\
        E_{q} &= \frac{(\omega - m_{1}\alpha_{1} - m_{2}\alpha_{2} - m_{3}\alpha_{3})^{2}}{4(u_{0}^{2} - u_{2}^{2})} - \frac{\left(\beta_{1} + (u_{3}^{2} + u_{2}^{2} + u_{1}^{2} - u_{0}^{2})\mu^{2}\right)}{4(u_{0}^{2} - u_{2}^{2})} \label{eq:Eq_radial}\\
        &\quad - \frac{u_{3}^{2} + u_{2}^{2} + u_{1}^{2} - 3u_{0}^{2}}{u_{0}^{2} - u_{2}^{2}}\,, \notag
    \end{align}
\end{subequations}
where $A_{3}$ is computed from \eqref{eq:coeffs_rad}, whereas $\theta_{u_{3}}$ is defined by Eq.~\eqref{eq:thetas}, and $\varPsi_{3}$ and $\chi_{3}$ are defined by Eq.~\eqref{eq:varPsis}. The conformal moduli $\lbrace t, q \rbrace$ and the accessory parameters $\lbrace E_{t}, E_{q} \rbrace$ are given by Eqs.~\eqref{eq:moduli_rad}  and \eqref{eq:Et_radial}--\eqref{eq:Eq_radial}, respectively. Thus, the resulting radial equation ~\eqref{eq:fuchsian_rad} corresponds to a Fuchsian equation with five singular points written in normal form. 

\subsubsection{Angular system}
\label{sec:2.2.2}
The angular system is given by two coupled second-order ODEs of the form
\begin{subequations}\label{eq:angular_Yi}
\begin{equation}
    \begin{split}
        \frac{1}{y_{i}}\frac{d}{d y_{i}}\left(y_{i}X_{i}(y_{i})\frac{d Y_{i}}{d y_{i}}\right) &+ \biggl[-\frac{1}{X_{i}(y_{i})}\biggl(y_{i}^{4}\Psi_{0} + y_{i}^{2}\Psi_{1} + \Psi_{2} + \frac{1}{y_{i}^{2}}\Psi_{3}\biggr)^{2} \\
        &+ \frac{1}{C_{3}}\frac{\Psi_{3}^{2}}{y_{i}^{2}} - m^{2}y_{i}^{4} + b_{1}y_{i}^{2} - b_{2}\biggr]Y_{i}(y_{i}) = 0\,, \qquad i = 1,2,
    \end{split}
\end{equation}
where
\begin{equation}
		X_{i}(y_{i}) = \frac{(L^{2} - y_{i}^{2})(a_{1}^{2} - y_{i}^{2})(a_{2}^{2} - y_{i}^{2})(a_{3}^{2} - y_{i}^{2})}{L^{2}y_{i}^{2}}\,,
\end{equation}
\end{subequations}
and $\Psi_{i}$'s are given by \eqref{eq:psis}. Now let us introduce dimensionless angular coordinates
\begin{equation}
    v_{i} = \frac{y_{i}}{L}\,, \qquad X_{i}(y_{i}) = L^{4}\Delta_{i}(v_{i})\,, \qquad \Delta_{i}(v_{i}) = \frac{(v_{i}^{2} - 1)(v_{i}^{2} - \alpha_{1}^{2})(v_{i}^{2} - \alpha_{2}^{2})(v_{i}^{2} - \alpha_{3}^{2})}{v_{i}^{2}}\,,
\end{equation}
and rescale the physical parameters and separation constants using \eqref{eq:dim_vars_params} and \eqref{eq:dim_psis}, so the resulting angular equations
\begin{equation}
    \begin{split}
            \frac{d^{2}Y_{i}}{d v_{i}^{2}} + \biggl(\frac{1}{v_{i}} + \frac{\Delta_{i}^{\prime}(v_{i})}{\Delta_{i}(v_{i})}&\biggr)\frac{d Y_{i}}{d v_{i}} + \Biggl[-\frac{\left(v_{i}^{6}\varPsi_{0} + v_{i}^{4}\varPsi_{1} +v_{i}^{2}\varPsi_{2} + \varPsi_{3}\right)^{2}}{v_{i}^{4}\Delta_{i}(v_{i})^{2}}\\
            &\qquad\qquad + \frac{1}{\Delta_{i}(v_{i})}\left(v_{i}^{2}\beta_{1} - \beta_{2} - v_{i}^{4}\mu^{2} + \frac{\varPsi_{3}^{2}}{v_{i}^{2}\chi_{3}}\right)\Biggr]Y_{i}(v_{i}) = 0\,,
    \end{split}
\end{equation}
do not depend on the AdS radius $L$, and can be rewritten as 
\begin{subequations}
    \begin{equation}\label{eq:dim_angular_Yi}
        \begin{gathered}
            \frac{d^{2}Y_{i}}{d v_{i}^{2}} + \biggl(\frac{1}{v_{i}} + \frac{\Delta_{i}^{\prime}(v_{i})}{\Delta_{i}(v_{i})}\biggr)\frac{d Y_{i}}{d v_{i}} - \Biggl[\frac{2 F_{0}}{v_{i}^{2} - 1} +\frac{2\alpha_{1}F_{1}}{v_{i}^{2} - \alpha_{1}^{2}} + \frac{2\alpha_{2}F_{2}}{v_{i}^{2} - \alpha_{2}^{2}} +\frac{2\alpha_{3}F_{3}}{v_{i}^{2} - \alpha_{3}^{2}}\\ 
            +\frac{2 G_{0} (v_{i}^{2} + 1)}{(v_{i}^{2} - 1)^{2}} + \frac{2 G_{1} (v_{i}^{2} +\alpha_{1}^{2} )}{(v_{i}^{2} - \alpha_{1}^{2})^{2}} + \frac{2 G_{2} (v_{i}^{2} + \alpha_{2}^{2})}{(v_{i}^{2} - \alpha_{2}^{2})^{2}} + \frac{2 G_{3} (v_{i}^{2} + \alpha_{3}^{2})}{(v_{i}^{2} - \alpha_{3}^{2})^{2}}\\
            - \frac{1}{\Delta_{i}(v_{i})}\left(v_{i}^{2}\beta_{1} - \beta_{2} - v_{i}^{4}\mu^{2} + \frac{\varPsi_{3}^{2}}{v_{i}^{2}\chi_{3}}\right)\Biggr]Y_{i}(v_{i}) = 0\,,
        \end{gathered}
    \end{equation}
where
    \begin{equation}\label{eq:coeffs_angular_Yi}
        \begin{split}
        &F_{k} = \lim_{v_{i} \to \alpha_{k}}\frac{d}{d v_{i}}\left[\frac{(v_{i} - \alpha_{k})^{2}(v_{i}^{6}\varPsi_{0} + v_{i}^{4}\varPsi_{1} + v_{i}^{2}\varPsi_{2} + \varPsi_{3})^{2}}{(v_{i}^{2} - \alpha_{0}^{2})^{2}(v_{i}^{2} - \alpha_{1}^{2})^{2}(v_{i}^{2} - \alpha_{2}^{2})^{2}(v_{i}^{2} - \alpha_{3}^{2})^{2}}\right]\,,\\
        &G_{k} = \lim_{v_{i} \to \alpha_{k}}\frac{(v_{i} - \alpha_{k} )^{2}(v_{i}^{6}\varPsi_{0} + v_{i}^{4}\varPsi_{1} + v_{i}^{2}\varPsi_{2} + \varPsi_{3})^{2}}{(v_{i}^{2} - \alpha_{0}^{2})^{2}(v_{i}^{2} - \alpha_{1}^{2})^{2}(v_{i}^{2} - \alpha_{2}^{2})^{2}(v_{i}^{2} - \alpha_{3}^{2})^{2}}\,, \qquad k = 0,1,2,3,
        \end{split}
    \end{equation}
\end{subequations}
and, $\alpha_{0} = 1$ for a consistent notation. It turns out that by substituting \eqref{eq:dim_psis} into \eqref{eq:coeffs_angular_Yi}, we find that
\begin{equation}
    \begin{split}
        &F_{0} = \frac{m_{1}\alpha_{1}\omega}{\alpha_{1}^{2} - 1} + \frac{m_{2}\alpha_{2}\omega}{\alpha_{2}^{2} - 1} + \frac{m_{3}\alpha_{3}\omega}{\alpha_{3}^{2} - 1} - \frac{\omega^{2}}{4}\,,\\
        &F_{j} = \sum_{i=1,i\neq j}^{3}\frac{m_{j}m_{i}\alpha_{i}}{\alpha_{j}^{2} - \alpha_{i}^{2}} + \frac{m_{j}\omega}{1 - \alpha_{j}^{2}} - \frac{m_{j}^{2}}{4\alpha_{j}}\,,\\
        &G_{0} = \frac{\omega^{2}}{4}\,, \qquad G_{j} = \frac{m_{j}^{2}}{4}\,, \qquad j = 1,2,3,
    \end{split}
\end{equation}
are invariant under $\left(\omega,m_{1},m_{2},m_{3}\right) \to \left(-\omega,-m_{1},-m_{2},-m_{3}\right)$. As expected, Eq.~\eqref{eq:dim_angular_Yi} also contains five regular singular points in the $v_{i}^{2}$ variable, located at $\alpha_{1}^{2}, \alpha_{2}^{2}, \alpha_{3}^{2}, 1, \infty$. Furthermore, the characteristic exponents of the Frobenius solutions near to each singularity are
\begin{equation}\label{eq:frob_Y}
    \begin{gathered}
        \rho^{\pm}_{\alpha_{1}} = \pm \frac{\vert m_{1} \vert}{2}\,, \quad
        \rho^{\pm}_{\alpha_{2}} = \pm \frac{\vert m_{2} \vert}{2}\,, \quad
        \rho^{\pm}_{\alpha_{3}} = \pm \frac{\vert m_{3} \vert}{2}\,, \quad
        \rho^{\pm}_{\alpha_{0}} = \pm \frac{\omega}{2}\,, \quad\\
        \rho^{\pm}_{\infty} = \frac{1}{2}\left(3 \pm \sqrt{9 + \mu^{2}}\right)\,.
    \end{gathered}
\end{equation}
Since the angular functions obey the same ODE, we will drop the subindex in $v_{i}$ and $Y_{i}$. In what follows, we will consider two different M\"{o}bius transformations to explore the space of parameters of the black hole. The first change of variables is
\begin{equation}\label{eq:zeta}
    \zeta = \frac{v^{2} - \alpha_{1}^{2}}{1 - \alpha_{1}^{2}}\,, 
\end{equation}
and the singularities are mapped to the Riemann sphere as
\begin{subequations}
    \begin{equation}
         \left( \alpha_{1}^{2}, \alpha_{2}^{2}, \alpha_{3}^{2}, 1, \infty \right) \, \longmapsto \, \left( 0, t, q, 1, \infty \right)
    \end{equation}
where
    \begin{equation}\label{eq:moduli_small}
        t = \frac{\alpha_{2}^{2} - \alpha_{1}^{2}}{1 - \alpha_{1}^{2}} \,, \qquad q = \frac{\alpha_{3}^{2} - \alpha_{1}^{2}}{1 - \alpha_{1}^{2}}\,.
    \end{equation}
\end{subequations}
It turns out that the hierarchy between the rotation parameters $0 < \alpha_{1} < \alpha_{2} < \alpha_{3}$ implies $0 < t < q < 1$. In addition, we define an $s$-homotopic transformation
\begin{equation}\label{eq:shomo_zeta}
    Y(\zeta) = \zeta^{-1/2}(\zeta - t)^{-1/2}(\zeta - q)^{-1/2}(\zeta-1)^{-1/2}\mathcal{Y}(\zeta)
\end{equation}
to bring Eq.~\eqref{eq:dim_angular_Yi} into an equation for $\mathcal{Y}(\zeta)$ of the form
\begin{subequations}
    \begin{equation}\label{eq:fuchsian_zeta}
	   \begin{split}
		\frac{d^{2}\mathcal{Y}}{d\zeta^{2}} + \biggl[\frac{\delta_{0}}{\zeta^{2}} &+ \frac{\delta_{t}}{(\zeta - t)^{2}} + \frac{\delta_{q}}{(\zeta - q)^{2}} + \frac{\delta_{1}}{(\zeta - 1)^{2}}  + \frac{\delta_{\infty} - \delta_{0} - \delta_{t} - \delta_{1} - \delta_{q}}{\zeta(\zeta - 1)}\\
		&\qquad\qquad + \frac{(t-1)E_{t}}{\zeta(\zeta - t)(\zeta - 1)} + \frac{(q-1)E_{q}}{\zeta(\zeta - q)(\zeta - 1)}\biggr]\mathcal{Y}(\zeta) = 0,
	   \end{split}
    \end{equation}
where
    \begin{equation}\label{eq:delta_small}
        \begin{gathered}
	        \delta_{0} = \frac{1}{4}\left(1 - m_{1}^{2}\right)\,, \quad \delta_{t} = \frac{1}{4}\left(1 - m_{2}^{2}\right)\,, \quad \delta_{q} = \frac{1}{4}\left(1 - m_{3}^{2}\right)\,, \quad \delta_{1} = \frac{1}{4}\left(1 - \omega^{2}\right)\,,\\
            \delta_{\infty} = \frac{1}{4}\left(1 - (9 + \mu^{2}) \right)\,,
        \end{gathered}
    \end{equation}
    \begin{equation}\label{eq:Et_angular}
        \begin{split}
            E_{t} &=\frac{(\alpha_{2}^{2} - \alpha_{1}^{2})m_{2}^{2}}{4\alpha_{2}^{2}} - \frac{\alpha_{1}m_{1}m_{2}}{2\alpha_{2}} - \frac{(\alpha_{2}^{2} - \alpha_{1}^{2})}{2\alpha_{2}}\left[\frac{\alpha_{3}m_{3}m_{2}}{\alpha_{2}^{2} - \alpha_{3}^{2}} + \frac{m_{2}\omega}{1 - \alpha_{2}^{2}}\right] \\
            &+ \frac{\left(\alpha_{2}^{4}\beta_{1} - \alpha_{2}^{2}\beta_{2} - \alpha_{2}^{6}\mu^{2} + \frac{\varPsi_{3}^{2}}{\chi_{3}}\right)}{4\alpha_{2}^{2}(1 - \alpha_{2}^{2})(\alpha_{3}^{2} - \alpha_{2}^{2})} + \frac{(\alpha_{2}^{2} - \alpha_{1}^{2})}{2(1 - \alpha_{2}^{2})} + \frac{(\alpha_{2}^{2} - \alpha_{1}^{2})}{2(\alpha_{3}^{2} - \alpha_{2}^{2})} - \frac{1}{2}\,,
        \end{split}
    \end{equation}
    \begin{equation}\label{eq:Eq_angular}
        \begin{split}
            E_{q} &= \frac{(\alpha_{3}^{2} - \alpha_{1}^{2})m_{3}^{2}}{4\alpha_{3}^{2}} - \frac{\alpha_{1}m_{1}m_{3}}{2\alpha_{3}} - \frac{(\alpha_{3}^{2} - \alpha_{1}^{2})}{2\alpha_{3}}\left[\frac{\alpha_{2}m_{2}m_{3}}{\alpha_{3}^{2} - \alpha_{2}^{2}} + \frac{m_{3}\omega}{1 - \alpha_{3}^{2}}\right]\\
            &- \frac{\left(\alpha_{3}^{4}\beta_{1} - \alpha_{3}^{2}\beta_{2} - \alpha_{3}^{6}\mu^{2} + \frac{\varPsi_{3}^{2}}{\chi_{3}}\right)}{4\alpha_{3}^{2}(1 - \alpha_{3}^{2})(\alpha_{3}^{2} - \alpha_{2}^{2})} + \frac{(1 - \alpha_{1}^{2})}{2(1-\alpha_{3}^{2})} - \frac{(\alpha_{2}^{2}-\alpha_{1}^{2})}{2(\alpha_{3}^{2}-\alpha_{2}^{2})} - \frac{3}{2}\,.
        \end{split}
    \end{equation}
\end{subequations}
As expected, the resulting angular equation~\eqref{eq:fuchsian_zeta} corresponds to a Fuchsian equation with five singular points written in normal form with conformal moduli $\lbrace t, q \rbrace$ and accessory parameters $\lbrace E_{t}, E_{q} \rbrace$ given by Eqs.~\eqref{eq:moduli_small} and \eqref{eq:Et_angular}--\eqref{eq:Eq_angular}, respectively. 

\section{Hill determinant and Accessory Parameter Expansions}
\label{sec:3}

Consider the normal form of a second-order linear ODE with five regular singular points 
\begin{equation}\label{eq:fuchsian_five}
	\begin{gathered}
		\frac{d^{2}\psi}{dz^{2}} + \Biggl[\frac{\delta_{0}}{z^{2}} + \frac{\delta_{t}}{(z-t)^{2}} + \frac{\delta_{1}}{(z-1)^{2}} + \frac{\delta_{q}}{(z-q)^{2}} + \frac{\delta_{\infty} - \delta_{0} - \delta_{t} - \delta_{1} - \delta_{q}}{z(z-1)}\\
		\qquad\qquad + \frac{(t-1)E_{t}}{z(z-t)(z-1)} + \frac{(q-1)E_{q}}{z(z-1)(z-q)}\Biggr]\psi(z) = 0\,,
	\end{gathered}
\end{equation}
where $\delta_{i} = \frac{1}{4} - a_{i}^{2}$ for $i=0,t,q,1,\infty$ are the semiclassical conformal weights directly related to the local monodromy exponents $\tfrac{1}{2} \pm a_{i}$ at the singular points $0,t,q,1,\infty$ and assume that $0 < \vert t \vert < \vert q \vert < 1$; $t$ and $q$ correspond to the conformal moduli, and the accessory parameters are $E_{t}$ and $E_{q}$.

Consider Floquet-type solutions of the form
\begin{equation}\label{eq:floquet_sigma}
        \psi_{\sigma,\pm}(z) = z^{\frac{1}{2}\pm\sigma}\sum_{n \in \mathbb{Z}}c_{n}z^{n}\,, 
\end{equation}
where the Laurent series converges to an analytic function in an annulus $\vert t \vert < \vert z \vert < \vert q \vert$. Similarly, we have
\begin{equation}\label{eq:floquet_rho}
        \psi_{\rho,\pm}(z) = z^{\frac{1}{2}\pm\rho}\sum_{n \in \mathbb{Z}}d_{n}z^{n}\,,
\end{equation}
for an annulus $\vert q \vert < \vert z \vert < 1$. Furthermore, $\sigma$ and $\rho$ are the Floquet exponents associated with the monodromies around the loop encircling two singular points $0$ and $t$, and three singular points $0$, $t$, and $q$, respectively.

\begin{figure}[!h]
\begin{center}
    \begin{tikzpicture}[scale=2]
        \begin{scope}[xshift=-2cm]

            \fill (0,0) circle (1pt) node[below] {$0$};
            \fill (1,0) circle (1pt) node[below] {$t$};

            \draw[thick, ->] (0.25,0)
            arc[start angle=0, end angle=360, radius=0.25];
            \node at (-0.25,0.3) {$\gamma_{0}$};

            \draw[thick, ->] (1.25,0)
            arc[start angle=0, end angle=360, radius=0.25];
            \node at (0.75,0.3) {$\gamma_{t}$};

            \draw[thick, ->] (0,1.3)
            arc[start angle=90, end angle=-270, radius=1.3];
            \node at (-0.9,1.15) {$\gamma_{\sigma}$};


        \end{scope}

        \begin{scope}[xshift=2cm]

            \fill (1,0) circle (1pt) node[below] {$q$};

            \draw[thick, ->] (0.25,0)
            arc[start angle=0, end angle=360, radius=0.25];
            \node at (-0.25,0.3) {$\gamma_{\sigma}$};

            \draw[thick, ->] (1.25,0)
            arc[start angle=0, end angle=360, radius=0.25];
            \node at (0.75,0.3) {$\gamma_{q}$};

            \draw[thick, ->] (0,1.3)
            arc[start angle=90, end angle=-270, radius=1.3];
            \node at (-0.9,1.15) {$\gamma_{\rho}$};


        \end{scope}
\end{tikzpicture}
\end{center}
\caption{The left figure describes the composite monodromy encompassing two singular points $0,t$ and their local monodromies. The right figure shows the monodromy around the loop encircling three singular points $0,t,q$.}
\label{fig:monodromies}
\end{figure}
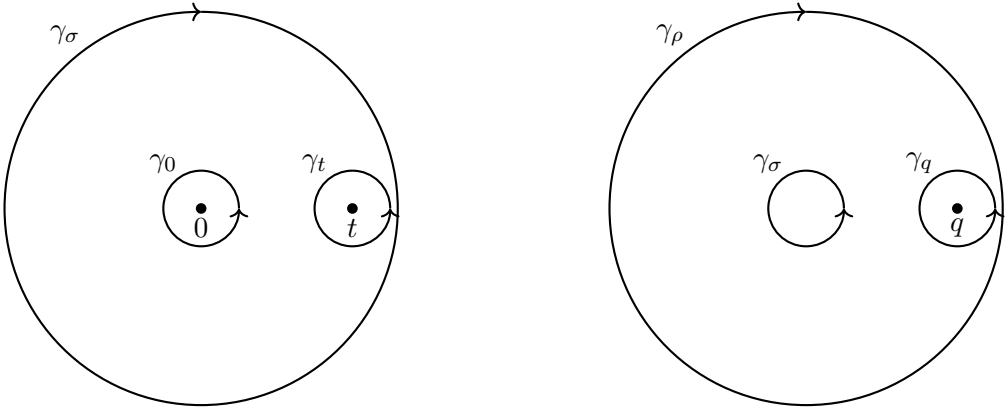

Then, we introduce series expansions of the accessory parameters
\begin{equation}\label{eq:expansions_small}
    E_{t} = \sum_{i,j = 0}^{\infty}\epsilon_{i,j}\left(\frac{t}{q}\right)^{i}q^{j}\,, \qquad E_{q} = \sum_{i,j = 0}^{\infty}\varepsilon_{i,j}\left(\frac{t}{q}\right)^{i}q^{j}\,,
\end{equation}
such that by substituting these expansions into the Fuchsian equation~\eqref{eq:fuchsian_five}, one can find the coefficients of \eqref{eq:expansions_small} from a system of linear equations that can be reformulated as a matrix equation which admits non--trivial solutions if and only if the determinant of the matrix vanishes, i.e., the Hill determinant is zero~\cite{wang1989special}\footnote{We thank Oleg Lisovyy for kindly sharing the implementation of the Hill determinant method.}. The first order expansions for small $t$ and $q$ in the regime $0 < \vert t \vert < \vert q \vert < 1$ are
\begin{subequations}\label{eq:ape_small}
    \begin{align}
            &E_{t} = \frac{1}{4} - \delta_{0} - \delta_{t} - \sigma^{2} + \frac{(\delta_{q} + \rho^{2} - \sigma^{2})(1 - 4\delta_{0} + 4\delta_{t} - 4\sigma^{2})}{2(1-4\sigma^{2})}\frac{t}{q} + \mathcal{O}\left(t^{2}q^{-2},t,q^{2}\right) \label{eq:Et_ape_small} \\
            &E_{q} = \sigma^{2} - \delta_{q} - \rho^{2} - \frac{(\delta_{q} + \rho^{2} - \sigma^{2})(1 - 4\delta_{0} + 4\delta_{t} - 4\sigma^{2})}{2(1-4\sigma^{2})}\frac{t}{q} \notag \\
            &\qquad\qquad\qquad\qquad + \frac{(1 + 4\delta_{1} - 4\delta_{\infty} - 4\rho^{2})(\delta_{q} - \rho^{2} + \sigma^{2})}{2-8\rho^{2}}q + \mathcal{O}\left(t^{2}q^{-2},t,q^{2}\right) \label{eq:Eq_ape_small}
    \end{align}
\end{subequations}
and higher order coefficients can be found in Appendix~\ref{sec:A.1}. These expansions reproduce the results in \cite{Liu:2024eut}.

On the other hand, we examine the regime $0 < \vert t \vert < 1 < \vert q \vert < \infty$, where the series expansions of the accessory parameters for small $t$ and large $q$ can be written as 
\begin{equation}\label{eq:expansions_large}
    	E_{t} = \sum_{i,j = 0}^{\infty}\tilde{\epsilon}_{i,j}t^{i}q^{-j}\,, \qquad E_{q} = \sum_{i,j = 0}^{\infty}\tilde{\varepsilon}_{i,j}t^{i}q^{-j}\,,
\end{equation}
which lead to the following series expansions
\begin{subequations}\label{eq:ape_large}
    \begin{align}
            &E_{t} = \frac{1}{4} - \delta_{0} - \delta_{t} - \sigma^{2} + \frac{(\delta_{1}+\rho^{2}-\sigma^{2})(1-4\delta_{0}+4\delta_{t}-4\sigma^{2})}{2(1-4\sigma^{2})}t + \mathcal{O}\left(t^{2},t q^{-1},q^{-2}\right)\,, \label{eq:Et_ape_large} \\
            &E_{q} = -\frac{1}{4} + \delta_{\infty} - \delta_{q} + \rho^{2} - \frac{(1 -4\delta_{\infty} + 4\delta_{q} - 4\rho^{2})(\delta_{1} - \rho^{2} + \sigma^{2})}{2(1 - 4\rho^{2})}\frac{1}{q} + \mathcal{O}\left(t^{2},t q^{-1},q^{-2}\right)\,, \label{eq:Eq_ape_large}
    \end{align}
\end{subequations}
and agree with the accessory parameter expansions computed from the instanton part of the Nekrasov--Shatashvili function in~\cite{Arnaudo:2025kof}. The two accessory parameter expansions are parametrized in terms of the Floquet exponents $(\sigma, \rho)$, which can be computed by inverting the Zamolodchikov relation between the accessory parameters and the semiclassical conformal blocks \cite{Lisovyy:2022flm}.

\subsection{Angular eigenvalues}
\label{sec:3.1}
We now address the computation of the separation constants in the slow-rotation limit, $0 < \alpha_{1} < \alpha_{2} < \alpha_{3} \ll 1$, which translates into the conformal moduli~\eqref{eq:moduli_small} as
\begin{equation}
    t = \alpha_{2}^{2}-\alpha_{1}^{2} + \mathcal{O}\left(\alpha_{1}^{2}\alpha_{2}^{2},\alpha_{1}^{4}\right)\,, \qquad q = \alpha_{3}^{2}-\alpha_{1}^{2} + \mathcal{O}\left(\alpha_{1}^{2}\alpha_{3}^{2},\alpha_{1}^{4}\right)\,,
\end{equation}
so that $0 < t < q < 1$.

In this regime, the separation constants $\beta_{1}$ and $\beta_{2}$ are obtained from equating the accessory parameter expansions~\eqref{eq:Et_ape_small} and \eqref{eq:Eq_ape_small} with the accessory parameters from the angular ODE \eqref{eq:Et_angular} and \eqref{eq:Eq_angular}, respectively. Introducing the dimensionless ratios
\begin{equation}\label{eq:angular_parametrization}
    \delta_{13} = \frac{\alpha_{1}}{\alpha_{3}}\,, \quad \delta_{23} = \frac{\alpha_{2}}{\alpha_{3}}\,,
\end{equation}
and performing a perturbative expansion in the regime $0 < \delta_{13} < \delta_{23} \ll \alpha_{3} \ll 1$, the separation constants are given by
\begin{subequations}\label{eq:beta_small_ser}
    \begin{align}
        &\beta_{1} = -4\left(1 - \rho^{2}\right) - 2\omega m_{3}\alpha_{3} - 2\omega\left(m_{1}\delta_{13} + m_{2}\delta_{23}\right)\alpha_{3} + 2m_{1}m_{2}\delta_{13}\delta_{23}\alpha_{3}^{2} + 2m_{1}m_{3}\delta_{13}\alpha_{3}^{2} \notag\\
        &\quad + 2m_{2}m_{3}\delta_{23}\alpha_{3}^{2} + \frac{1}{2}\left[\omega^{2} + \mu^{2} + m_{3}^{2} - 4\rho^{2} + 4\sigma^{2} - \frac{(m_{3}^{2}-4\sigma^{2})(\omega^{2} - (9+\mu^{2}))}{1-4\rho^{2}}\right]\alpha_{3}^{2} \notag\\ 
        &\quad + \beta_{1,2,0,2}\delta_{13}^{2}\alpha_{3}^{2} 
        + \beta_{1,0,2,2}\delta_{23}^{2}\alpha_{3}^{2} + \beta_{1,0,0,4}\alpha_{3}^{4} + \ldots\,, \label{eq:beta_one} \\[8pt]
        &\beta_{2} = -\left(1-4\sigma^{2}\right)\alpha_{3}^{2} + 2m_{1}m_{3}\delta_{13}\alpha_{3}^{2} + 2m_{2}m_{3}\delta_{23}\alpha_{3}^{2} + 2m_{1}m_{2}\delta_{13}\delta_{23}\alpha_{3}^{2} -2\omega m_{1}\delta_{13}\alpha_{3}^{3} \notag\\ 
        &\quad -2\omega m_{2}\delta_{23}\alpha_{3}^{3} + \beta_{2,2,0,2}\delta_{13}^{2}\alpha_{3}^{2} + \beta_{2,0,2,2}\delta_{23}^{2}\alpha_{3}^{2} \ldots\,, \label{eq:beta_two}
    \end{align}
\end{subequations}
where the coefficients $\beta_{m,i,j,k}$ for $m=1,2$, are listed in Appendix~\ref{sec:A.2}. In Appendix~\ref{sec:B}, we derive the separation constants for the case $0 < t < 1 < q < \infty$. According to Eq.~\eqref{eq:moduli_large}, this corresponds to $\alpha_{1} < \alpha_{2} < \alpha_{3}$, in which the two largest rotation parameters $\alpha_{2},\alpha_{3} \to 1$, while $\alpha_{1}$ remains arbitrary.

In order to fully determine the separation constants, we need to impose boundary conditions on the solutions of the angular equation \eqref{eq:fuchsian_zeta}. Since there are two Floquet exponents $(\sigma,\rho)$, we require that solutions of $\mathcal{Y}(\zeta)$ are regular in the domains $\zeta \in \left[0, t\right]$ and $\zeta \in \left[t, q\right]$\footnote{In general, we impose regularity of the solutions for $Y_{1}$ in the domain $v_{1}^{2} \in \left[\alpha_{1}^{2}, \alpha_{2}^{2}\right]$ and of the solutions for $Y_{2}$ in the domain $v_{2}^{2} \in \left[\alpha_{2}^{2}, \alpha_{3}^{2}\right]$. Nevertheless, both angular functions satisfy the same ODE.}. By inspecting the indicial coefficients of the Frobenius solutions around the involved singular points, we find that the difference between them is an integer. Thus, the second local solution develops a logarithmic term at each singular point. More precisely, 
\begin{equation}
    \begin{cases}
        \mathcal{Y}^{(k)}_{\rm reg} \sim (\zeta - \zeta_{k})^{\frac{1}{2}(1 + \vert m_{k}\vert)}\,, \\
        \\
        \mathcal{Y}^{(k)}_{\rm irr} \sim (\zeta - \zeta_{k})^{\frac{1}{2}(1 - \vert m_{k}\vert)} + A_{\zeta_{k}}(\zeta - \zeta_{k})^{\frac{1}{2}(1 + \vert m_{k}\vert)}\log(\zeta - \zeta_{k})\,,
    \end{cases}
    \quad \zeta \to \zeta_{k} = 0,t,q\,,
\end{equation}
for some coefficients $A_{\zeta_{k}}$ and $m_{k}\,(k=1,2,3) \in \mathbb{Z}$. In the basis $\left(\mathcal{Y}^{(k)}_{\rm reg}, \, \mathcal{Y}^{(k)}_{\rm irr}\right)$, the logarithmic structure of these solutions implies that the local monodromy matrices take the Jordan form
\begin{equation}
    J_{k} = e^{i\pi(1+\vert m_{k} \vert)}
    \begin{pmatrix}
        1& 1\\
        0& 1
    \end{pmatrix}\,.
\end{equation}
For Fuchsian systems with four singular points at $0,t,1\infty$ and one non-logarithmic singularity at $\zeta = q$, the vanishing condition for $A_{q}$ implies that $q(t)$ satisfies the Painlev\'{e} VI equation \cite{Fuchs1907}. The so-called apparent singularity condition was interpreted in \cite{Jia:2024zes} as the insertion of degenerate primary operators, while the connection coefficients associated with local solutions exhibiting logarithmic tails were computed in \cite{He:2023wcs,Ren:2024hwf}. 
 
Following the procedure described in \cite{Novaes:2018fry}, we consider two linearly independent solutions of the angular equation \eqref{eq:fuchsian_zeta}, whose local behavior near $\zeta=0$ and $\zeta=t$ is given by
\begin{equation}\label{eq:ang_solution_at_t}
    \mathcal{Y}^{(\rm reg)} \sim 
    \begin{cases}
        A^{(\sigma)}_{\rm reg}\mathcal{Y}^{(t)}_{\rm reg} + A^{(\sigma)}_{\rm irr}\mathcal{Y}^{(t)}_{\rm irr}\,,  &\zeta \to t\,,\\
        \\
        \mathcal{Y}^{(0)}_{\rm reg}\,,  &\zeta \to 0\,,
    \end{cases}
\end{equation}
and
\begin{equation}
    \mathcal{Y}^{(\rm irr)} \sim
    \begin{cases}
        \tilde{A}^{(\sigma)}_{\rm irr}\mathcal{Y}^{(t)}_{\rm irr} + \tilde{A}^{(\sigma)}_{\rm reg}\mathcal{Y}^{(t)}_{\rm reg}\,,  &\zeta \to t\,,\\
        \\
        \mathcal{Y}^{(0)}_{\rm irr}\,,  &\zeta \to 0\,.
    \end{cases}
\end{equation}
More precisely, we choose a particular solution near $\zeta=0$ and express it in the basis of local solutions near $\zeta=t$. Here, $A^{(\sigma)}_{\rm reg}$, $A^{(\sigma)}_{\rm irr}$, $\tilde{A}^{(\sigma)}_{\rm reg}$, $\tilde{A}^{(\sigma)}_{\rm irr}$ denote the corresponding connection coefficients. 

We further assume that the Floquet exponents parametrize the composite monodromies associated with the cycles shown in Fig.~\ref{fig:monodromies}. A fundamental matrix of solutions can then be written as $\left( \mathcal{Y}^{(\rm reg)}, \, \mathcal{Y}^{(\rm irr)}\right) = \mathcal{Y}^{(k)}g_{k}$, where the corresponding monodromy matrices are $M_{k} = g_{k}^{-1}J_{k}g_{k}$, with $k = 0,t$. It turns out that the composite monodromy associated with the contour encircling the singularities at $\zeta = 0$ and $\zeta = t$ is characterized by the Floquet exponent $\sigma$ through
\begin{equation}
    \mathrm{Tr}\,M_{t}M_{0} = - 2\cos 2\pi\sigma\,,
\end{equation}
and the computation of $\sigma$ is therefore reduced to a connection problem, with the corresponding connection coefficient given by
\begin{equation}
    A^{(\sigma)}_{\rm irr} = \pm\sqrt{2(1+e^{i\pi(\vert m_{1}\vert + \vert m_{2}\vert)}\cos(2\pi\sigma))}\,.
\end{equation}
The regularity of the local solutions is ensured by imposing $A^{(\sigma)}_{\rm irr} = 0$, which yields the angular quantization condition
\begin{equation}\label{eq:sigma}
    \sigma = \frac{1}{2}\left(\vert m_{1}\vert + \vert m_{2}\vert + 2j +1 \right)\,, \qquad j \in \mathbb{Z}\,.
\end{equation}
The same construction applies to the composite monodromy around the three singular points $0$, $t$, and $q$, which is parametrized by the second Floquet exponent $\rho$ as
\begin{equation}
    \mathrm{Tr}\,M_{q}M_{\sigma} = - 2\cos 2\pi\rho\,.
\end{equation}
Analogously, we denote by $A^{(\rho)}_{\rm reg}, A^{(\rho)}_{\rm irr}$ and their tilded counterparts as the corresponding connection coefficients. A straightforward computation gives
\begin{equation}
    A^{(\rho)}_{\rm irr} = \pm\sqrt{2(1+e^{-i\pi(2\sigma + \vert m_{3}\vert)}\cos(2\pi\rho))}\,,
\end{equation}
and the regularity condition again requires its vanishing, leading to the second angular quantization condition 
\begin{equation}\label{eq:rho}
    \rho = \frac{1}{2}\left(\vert m_{1}\vert + \vert m_{2}\vert + \vert m_{3}\vert + 2j + 2k + 2\right)\,, \qquad j,k \in \mathbb{Z}\,.
\end{equation}
Substituting the quantization conditions \eqref{eq:sigma} and \eqref{eq:rho} into Eq. \eqref{eq:beta_small_ser} yields perturbative expansions of the angular eigenvalues analogous to the small $a_{1}, a_{2}$ expansions derived in \cite{Cho:2011yp} for $D \geq 6$ dimensional Kerr--(A)dS with only two independent rotation parameters and all the others set equal to zero. 

In the non-rotating limit, the separation constants should recover the eigenvalues of the Laplace--Beltrami operator on the unit five--sphere $\Delta_{S^{5}} Y_{\ell}^{m_{1},m_{2},m_{3}} = -\ell(\ell+4)Y^{m_{1},m_{2},m_{3}}_{\ell}$, where $\ell \geq 0$. It turns out that $\beta_{2}$ vanishes and $\beta_{1}$ reduces to the constant term in Eq.~\eqref{eq:beta_one}. Then, we find 
\begin{equation}
    \rho = \pm \frac{\ell+2}{2}\,,
\end{equation}
where $\ell$ is the total angular momentum quantum number, which is equivalent to
\begin{equation}\label{eq:ang_eigen}
    \ell = 2(j + k) + \vert m_{1}\vert + \vert m_{2}\vert + \vert m_{3}\vert\,, \quad j,k \in \mathbb{Z}
\end{equation}
and the assumption $j + k \geq 0$ yields the constraint
\begin{equation}
    \ell \geq \vert m_{1}\vert + \vert m_{2}\vert + \vert m_{3}\vert\,.
\end{equation}
Recently, the connection coefficients and local solutions of second order linear ODEs of Fuchsian type have been computed using conformal blocks and crossing symmetry \cite{Bonelli:2021uvf,Bonelli:2022ten,Lisovyy:2022flm}. In particular, ODEs with five regular singularities are obtained from the semiclassical limit of the corresponding BPZ equation such that Frobenius solutions correspond to the semiclassical limit of a particular five--point conformal block with the insertion of an extra degenerate field \cite{Jia:2024zes,Liu:2024eut,Arnaudo:2025kof}. These results can be exploited to express the local solutions around different singular points, together with the corresponding connection coefficients, in terms of the black hole parameters. In particular, the angular quantization conditions follow from imposing regularity of the local solutions entering the boundary conditions. The Floquet exponent $\sigma$ is determined by the connection problem between the singular points at $\zeta = 0$ and $\zeta = t$, whereas $\rho$ is obtained by the connection problem associated with $\zeta = t$ and $\zeta = q$. As we show below, the regularity conditions translate into the following quantization conditions
\begin{equation}
    \tfrac{1}{2} + a_{0} + \sigma + a_{t} = -j\,, \qquad \tfrac{1}{2} - \sigma + a_{q} -\rho = -k\,, \qquad j,k \in \mathbb{Z}_{\geq 0}
\end{equation}
which are in agreement with Eq.~\eqref{eq:ang_eigen}. Once the angular eigenvalues are found~\eqref{eq:beta_small_ser}, one can solve the radial eigenvalue problem associated with Eq.~\eqref{eq:fuchsian_rad}. In contrast to the angular eigenvalues, which are obtained by directly solving the accessory parameter equations, the computation of radial eigen-frequencies first requires determining the Floquet exponents and then using the connection coefficients to extract them.

\subsection{Quasinormal mode frequencies}
\label{sec:3.2}
We compute the quasinormal modes in the regime $0 < \alpha_{1} < \alpha_{2} < \alpha_{3} < u_{3} \ll 1$, by introducing the parametrization 
\begin{equation}
    \alpha_{1} = \gamma_{13}u_{3}\,, \quad \alpha_{2} = \gamma_{23}u_{3}\,, \quad \alpha_{3} = \gamma_{33}u_{3}\,,
\end{equation}
and assuming 
\begin{equation}
    0 < \gamma_{13} < \gamma_{23} < \gamma_{33} \ll u_{3} \ll 1\,.
\end{equation}
By expanding the radial conformal moduli~\eqref{eq:moduli_rad}, we have
\begin{equation}\label{eq:small_t_and_large_q}
    \begin{split}
        t &= -1 + \boldsymbol{\gamma}^{2} + 3u_{3}^{2} + \mathcal{O}\left(u_{3}^{4},u_{3}^{2}\boldsymbol{\gamma}^{2},\boldsymbol{\gamma}^{4}\right)\,,\\
        q &= -\frac{1-\boldsymbol{\gamma}^{2}+\mathcal{O}(\boldsymbol{\gamma}^{4})}{u_{3}^{2}} + 2 + u_{3}^{2} -\boldsymbol{\gamma}^{2} + \mathcal{O}\left(u_{3}^{4},u_{3}^{2}\boldsymbol{\gamma}^{2},\boldsymbol{\gamma}^{4}\right)\,,
    \end{split}
\end{equation}
where
\begin{equation}
    \boldsymbol{\gamma} = (\gamma_{13},\,\gamma_{23},\,\gamma_{33})\,, \qquad \boldsymbol{\gamma}^{2} \equiv \boldsymbol{\gamma}\cdot\boldsymbol{\gamma} = \gamma_{13}^{2} + \gamma_{23}^{2} + \gamma_{33}^{2}\,.
\end{equation}
Thus, the locations of the singular points of the Fuchsian ODE~\eqref{eq:fuchsian_rad} satisfy $0 < \vert t \vert < 1 < \vert q \vert < \infty$. The radial dictionary for $a_{i}$, using $\delta_{i} = \frac14 - a_{i}^{2}$ with $\delta_{i}$ defined in \eqref{eq:delta_radial}, can be expanded as follows
\begin{equation}\label{eq:radial_dictionary}
    \begin{aligned}
        &a_{0} = 0\,,\\
        &a_{t} = \frac{i}{4} \bigl(\omega u_{3} - m_{1}\gamma_{13} - m_{2}\gamma_{23} - m_{3}\gamma_{33}\bigr) + \mathcal{O}\left(u_{3}\boldsymbol{\gamma}^{2},u_{3}^{2}\boldsymbol{\gamma},u_{3}^{3},\boldsymbol{\gamma}^{3}\right)\,,\\
        &a_{1} = \frac14\bigl(\omega u_{3} + m_{1}\gamma_{13} + m_{2}\gamma_{23} + m_{3}\gamma_{33}\bigr) + \mathcal{O}\left(u_{3}^{3},\boldsymbol{\gamma}^{3}\right)\,,\\
        &a_{q} = \frac{\sqrt{9 + \mu^{2}}}{2}\,,\\
        &a_{\infty} = \frac{\omega}{2} + \mathcal{O}\left(u_{3}^{4}\right)\,.
    \end{aligned}
\end{equation}
Furthermore, to determine the Floquet exponents, we introduce the following ansatz for $\sigma$ and $\rho$:
\begin{equation}\label{eq:sigma_and_rho}
    \sigma = \sum_{i,j = 0}^{\infty}\varsigma_{i,j}t^{i}q^{-j}\,, \qquad \rho = \sum_{i,j = 0}^{\infty}\varrho_{i,j}t^{i}q^{-j}\,.
\end{equation}
Substituting these expansions into the accessory parameter expansions,  Eqs.~\eqref{eq:Et_ape_large} and \eqref{eq:Eq_ape_large}, the coefficients $\varsigma_{i,j}$ and $\varrho_{i,j}$ are then computed recursively order by order in terms of the parameters $\delta_{i}$, $E_{t}$, and $E_{q}$. The first coefficients are
\begin{align}
    \sigma &= \frac{1}{2}\sqrt{1-4E_{t}-4\delta_{0}-4\delta_{t}} + \frac{(E_{t}+E_{q}+\delta_{0}+\delta_{t}+\delta_{q}+\delta_{1}-\delta_{\infty})(E_{t}+2\delta_{t})}{2(E_{t}+\delta_{0}+\delta_{t})\sqrt{1-4E_{t}-4\delta_{0}-4\delta_{t}}}t \nonumber\\
    &\quad + \mathcal{O}\left(t^{2},t/q,q^{-2}\right)\,, \label{eq:sigma_ape}\\
    \rho &= \frac{1}{2}\sqrt{1+4E_{q}+4\delta_{q}-4\delta_{\infty}} -\frac{E_{q}(E_{q}+E_{t}+\delta_{0}+\delta_{t}+\delta_{q}-\delta_{1}-\delta_{\infty})}{2(E_{q}+\delta_{q}-\delta_{\infty})\sqrt{1+4E_{q}+4\delta_{q}-4\delta_{\infty}}}\frac{1}{q} \nonumber\\
    &\quad + \mathcal{O}\left(t^{2},t/q,q^{-2}\right)\,, \label{eq:rho_ape}
\end{align}
where $\varsigma_{i,j} = 0$ for $i < j$ and $\varrho_{i,j} = 0$
for $j < i$, and we use $\delta_{i} = \tfrac14 - a_{i}^{2}$, with $a_{i}$ defined in~\eqref{eq:radial_dictionary}. There is a subtlety in the computation of $\sigma$ arising from the fact that $\vert t \vert \sim \mathcal{O}(1)$, which implies that higher-order terms in the expansion can contribute at the same order as lower-order terms once they are re-expressed in terms of the black hole parameters\footnote{We are indebted to Paolo Arnaudo for clarifying this point.}. Nevertheles, the expansion is expected to have the structure $\sigma = \sigma_{0} + \sigma_{2}u_{3}^{2} + \mathcal{O}(u_{3}^{4})$.

By contrast, $\rho$ does not receive these contributions, so that substituting Eqs.~\eqref{eq:moduli_rad}, \eqref{eq:delta_radial}, and \eqref{eq:Eq_radial} into the Floquet exponent expansion \eqref{eq:rho_ape} and expanding perturbatively in the horizon radius and the rotation parameters yields
\begin{equation}\label{eq:rho_rad}
    \begin{split}
        \rho &= \frac{\ell+2}{2} + \Biggl[\frac{15\ell(\ell+2)(\ell+4)}{32(\ell+1)(\ell+3)}-\frac{\mu^{2}(\mu^{2}-6\omega^{2}+6\ell(\ell+4)+32)}{32(\ell+1)(\ell+2)(\ell+3)}\\
        &\qquad -\frac{5\omega^{2}(\omega^{2} - 6\ell(\ell+4)-28)}{32(\ell+1)(\ell+2)(\ell+3)}\Biggr]u_{3}^{4} + \ldots
    \end{split}
\end{equation}
The QNMs are solutions to the eigenvalue problem associated with \eqref{eq:fuchsian_rad} that satisfy the following boundary conditions: a purely ingoing wave at the event horizon $(z=t)$ and regularity at spatial infinity $(z=q)$. We therefore consider radial solutions with the following asymptotic behavior:
\begin{equation}\label{eq:boundary_rad}
    \psi(z) \sim
    \begin{cases}
        (z - t)^{\frac{1}{2}(1 - \theta_{u_{3}})}\,, &z \to t\,,\\
        \\
        \mathcal{C}_{q,-}(z - q)^{\frac{1}{2}(1 - \sqrt{9+\mu^{2}})} + \mathcal{C}_{q,+}(z - q)^{\frac{1}{2}(1 + \sqrt{9+\mu^{2}})}\,, &z \to q\,,
    \end{cases}
\end{equation}
where $\mathcal{C}_{q,\pm}$ are constants proportional to the connection coefficients relating the local Frobenius solutions around $z=t$ and $z=q$. For $\mu > 0$, at $z \to q$ the first solution diverges, whereas the second converges, and thus, these solutions will correspond to non-normalizable and normalizable solutions, respectively. Furthermore, the mass of the scalar field is related to the conformal dimension $\Delta$ of a dual CFT operator through
\begin{equation}
    \mu^{2} = \Delta(\Delta-6)\,.
\end{equation}
The connection formulas between Frobenius solutions around different singular points were derived in closed form using the AGT correspondence~\cite{Arnaudo:2025kof}. In this formalism,
the connection coefficients are parametrized by two parameters $b_{1}$ and $b_{2}$, which represent the v.e.v. of the scalars in the $\mathcal{N}=2$ vector multiplets and satisfy the Matone relations~\cite{Matone:1995rx}
\begin{equation}
    \begin{split}
        &E_{t} = -\frac14 - b_{1}^{2} + a_{0}^{2} + a_{t}^{2} + t\partial_{t}F(1/q,t)\,,\\
        &E_{q} = -\frac14 + b_{2}^{2} + a_{q}^{2} - a_{\infty}^{2} + q\partial_{q}F(1/q,t)\,.
    \end{split}
\end{equation}
and $F(1/q,t)$ is the instanton part of the Nekrasov--Shatashvili (NS) function of the four-dimensional $\mathcal{N}=2$ superconformal linear quiver gauge theory with gauge group $SU(2) \times SU(2)$. In the present work, however, the accessory parameter expansions are computed using the Hill determinant method ~\eqref{eq:ape_large}, where the natural variables are the Floquet exponents $\sigma$ and $\rho$. Since the accessory parameter expansions obtained from the Matone relations agree with those derived from the Hill determinant approach in the asymptotic regimes considered here, we identify
\begin{equation}
    b_{1} \equiv \sigma\,, \qquad b_{2} \equiv \rho\,,
\end{equation}
and adopt the notation $(\sigma,\rho)$ throughout the remainder of this work. With this identification, Frobenius solutions around $z=t$ and $z=q$ reads
\begin{equation}\label{eq:frobsol}
    \begin{aligned}
    &\psi_{t,\theta_{1}}(z) = t^{\theta_{1}a_{t}}e^{\frac{1}{2}\theta_{1}\partial_{a_{t}}F(1/q,t)}\\
    &\quad \times \sum_{\theta_{2},\theta_{3},\theta_{4}=\pm}e^{i\pi(\frac12 - \theta_{2}\sigma + \theta_{3}\rho +\theta_{4}a_{q})}
    \mathcal{M}_{\theta_{1}\theta_{2}}\left(a_{t},\sigma;a_{0}\right)\mathcal{M}_{(-\theta_{2})\theta_{3}}\left(\sigma,\rho;a_{1}\right)\mathcal{M}_{(-\theta_{3})\theta_{4}}\left(\rho,a_{q};a_{\infty}\right)\\
    &\quad \times\, t^{\theta_{2}\sigma}e^{-\frac{\theta_{2}}{2}\partial_{\sigma}F(1/q,t)-\frac{\theta_{3}}{2}\partial_{\rho}F(1/q,t)-\frac{\theta_{4}}{2}\partial_{a_{q}}F(1/q,t)}q^{-(\theta_{3}\rho + \theta_{4}a_{q})}\psi_{q,\theta_{4}}(z)\,,
    \end{aligned}
\end{equation}
and the connection matrix $\mathcal{M}_{\theta\theta^{\prime}}$ is given by
\begin{equation}\label{eq:matrix_M}
    \mathcal{M}_{\theta \theta'}\left(a_{1},a_{2};a_{3}\right) = \frac{\Gamma\left(-2\theta'a_{2}\right)\Gamma\left(1+2\theta a_{1}\right)}{\Gamma\left(\tfrac{1}{2} + \theta a_{1} - \theta'a_{2} + a_{3}\right)\Gamma\left(\tfrac{1}{2} + \theta a_{1} - \theta'a_{2} - a_{3}\right)}\,, \quad \theta,\theta'=\pm\,.
\end{equation}
According to the boundary conditions~\eqref{eq:boundary_rad}, the ingoing wave corresponds to the solution with $\theta_{1}=-1$, namely $\psi_{t,-}(z)$ in \eqref{eq:frobsol}. We then express $\psi_{t,-}(z)$ as a linear combination of $\psi_{q,\pm}(z)$. By imposing regularity at spatial infinity, we require that $\mathcal{C}_{q,-} = 0$, and hence $\theta_{4}=-1$. The resulting condition can be schematically written as
\begin{equation}\label{eq:cond_Cqminus}
    \mathcal{C}^{(+,+)}_{q,-}t^{\sigma}q^{-\rho} + \mathcal{C}^{(+,-)}_{q,-}t^{\sigma}q^{\rho} + \mathcal{C}^{(-,+)}_{q,-}t^{-\sigma}q^{-\rho} + \mathcal{C}^{(-,-)}_{q,-}t^{-\sigma}q^{\rho}= 0\,,
\end{equation}
where
\begin{equation}\label{eq:coef_Cqminus}
    \begin{gathered}
        \mathcal{C}^{(\theta_{2},\theta_{3})}_{q,-} = \mathcal{M}_{-\theta_{2}}(a_{t},\sigma;a_{0})\mathcal{M}_{(-\theta_{2})\theta_{3}}(\sigma,\rho;a_{1})\mathcal{M}_{(-\theta_{3})-}(\rho,a_{q};a_{\infty})\\
        \times e^{-i\pi(\theta_{2}\sigma - \theta_{3}\rho)-\frac{\theta_{2}}{2}\partial_{\sigma}F(1/q,t)-\frac{\theta_{3}}{2}\partial_{\rho}F(1/q,t)}\,.
    \end{gathered}
\end{equation}
Furthermore, using the black hole parametrization~\eqref{eq:small_t_and_large_q} and Eq.~\eqref{eq:rho_rad}, we get
\begin{equation}
    \quad q^{-\rho} \sim u_{3}^{\ell+2} \to 0\,,
\end{equation}
in the small black hole limit for any fixed $\ell$. Thus, the terms containing $q^{-2\rho}$ in Eq.~\eqref{eq:cond_Cqminus} are suppressed in the small-$u_{3}$ limit, yielding 
\begin{equation}\label{eq:cond_Cqminus_reduced}
    \mathcal{C}^{(+,-)}_{q,-}t^{2\sigma} + \mathcal{C}^{(-,-)}_{q,-} + \mathcal{O}\left(u_{3}^{2\ell+4}\right) = 0\,.
\end{equation}
Both coefficients contain the common factor $\mathcal{M}_{+-}(\rho,a_{q};a_{\infty})$, which can be factored out. Therefore, the radial quantization condition is given by
\begin{equation}\label{eq:M_pm}
    \mathcal{M}_{+-}(\rho,a_{q};a_{\infty}) = 0\,,
\end{equation}
and agrees with the result obtained in~\cite{Arnaudo:2025kof}. Then, the QNM frequencies can be determined by the poles of the Gamma functions in Eq.~\eqref{eq:M_pm}, and without loss of generality, we choose
\begin{equation}\label{eq:qnm_cond_large}
    \frac12 + \rho + a_{q} \pm a_{\infty} = -n\,, \qquad n \in \mathbb{Z}_{\geq 0}\,,
\end{equation}
where the different sign $a_{\infty}$ will be related to the sign of the real part of the QNM frequencies. Furthermore, we replace the radial dictionary \eqref{eq:radial_dictionary} into Eq.~\eqref{eq:qnm_cond_large} and consider the following expansion for the real part of the frequencies
\begin{equation}\label{eq:omega}
    \omega^{\rm (Re)}_{n,\ell} = \sum_{i,j,k,l = 0}\Omega_{i,j,k,l}\gamma_{13}^{i}\gamma_{23}^{j}\gamma_{33}^{k}u_{3}^{l}\,,
\end{equation}
so the first correction to the normal modes is of the form
\begin{equation}\label{eq:qnm_re_large_q}
    \begin{split}
        \omega^{\rm (Re)}_{n,\ell} = 2n+\ell+\Delta &- \Biggl[\frac{5}{2}(2n+\ell+\Delta) - \frac{15\ell(\ell+2)(\ell+4)}{16(\ell+1)(\ell+3)}\\ 
        &\quad + \frac{\Delta(\Delta-6)(\Delta(\Delta-6)-6(2n+\ell+\Delta)^{2}+6\ell(\ell+4)+32)}{16(\ell+1)(\ell+2)(\ell+3)}\\
        &\quad + \frac{5(2n+\ell+\Delta)^{2}((2n+\ell+\Delta)^{2} - 6\ell(\ell+4)-28)}{16(\ell+1)(\ell+2)(\ell+3)}\Biggr]u_{3}^{4} + \ldots\,,
    \end{split}
\end{equation}
and reproduces the expression for the real part of the quasinormal modes in Schwarzschild--AdS$_{7}$ derived in \cite{Arnaudo:2025kof,Wang:2014eha}. We have not found corrections due to the black hole rotations up to fifth total  order in the perturbative expansion of $\gamma_{13}$, $\gamma_{23}$, $\gamma_{33}$, and $u_{3}$. This suggests that such corrections first appear at  sixth total order. Since higher-order expansions are beyond the scope of our computation, we focus on the non-rotating limit of the eigenfrequencies.  

\section{Discussion}
\label{sec:4}
In this paper, we studied scalar perturbations of seven-dimensional Kerr--AdS black holes using the method of accessory parameter expansions. To this end, we employed the Hill determinant method to derive analytic expansions of the accessory parameters associated with a Fuchsian ODE with five singular points. These expansions are expressed in terms of the conformal moduli, the conformal weights of the associated Fuchsian equations, and the Floquet exponents $(\sigma,\rho)$. Furthermore, the resulting expansions are in agreement with those derived from the instanton part of the Nekrasov--Shatashvili function of four-dimensional $\mathcal{N}=2$ quiver gauge theories~\cite{Liu:2024eut,Arnaudo:2025kof}.

The accessory parameter expansions can be used directly to obtain perturbative expressions for the separation constants in the slowly rotating limit, thereby solving the angular eigenvalue problem. We also derived analytic expansions for the regime in which two rotation parameters $\alpha_{2}$ and $\alpha_{3}$ approach to 1, for fixed $\alpha_{1} \ll 1$, as discussed in Appendix~\ref{sec:B}. 

The computation of the QNM frequencies is, however, more subtle. In this case, the accessory parameter expansions must first be inverted to obtain perturbative expansions for the Floquet exponents. When the Floquet exponents are re-expressed in terms of the black hole parameters, formally higher order terms in the conformal moduli contribute at lower orders in the perturbative expansion. As a result, one of the Floquet exponents cannot be determined consistently order by order. Interestingly, the same feature occurs for different M\"{o}bius transformations, corresponding to distinct parameter regimes, suggesting that it is intrinsic to the asymptotically AdS black hole rather than to a particular parametrization of the Fuchsian equation. Nevertheless, the Floquet exponent entering the radial quantization condition remains unaffected and can therefore be used to determine the real part of the QNM frequencies. We obtain a perturbative expansion including terms up to fourth order in $u_{3}$. Our results indicate that the first corrections due to the rotation parameters appear at sixth total order in the perturbative expansion. 

The imaginary part of the QNM frequencies requires determining the poles of the retarded Green's function, which correspond to the vanishing of the connection coefficient associated with the non-normalizable solution. In particular, identifying these poles depends on the precise computation of the Floquet exponent $\sigma$, whose perturbative expansion remains to be fully established. 

The coalescence of two singular points reduces the Fuchsian ODE with five singularities to an equation with three singularities and one irregular singular point. This confluent limit has recently been studied as a decoupling limit of the $\mathcal{N}=2$ quiver gauge theory to investigate the angular spectrum of the extremal C--metric \cite{Yang:2026dpt}. A natural direction for future work is to analyze the extremal limit of the Kerr--AdS$_{7}$ black hole using the Hill determinant method, compare the resulting accessory parameter expansions with those obtained from the corresponding quiver gauge theory, and compute the quasinormal mode frequencies. 

\acknowledgments

J.B.A. acknowledges financial support from the Funda\c{c}\~{a}o para a Ci\^{e}ncia e a Tecnologia (FCT) through the research project UID/00208/2025 (DOI: 10.54499/UID/00208/2025). The author thanks Paolo Arnaudo, Bruno da Cunha, Jo\~ao Cavalcante, Alba Grassi, and Oleg Lisovyy for valuable discussions.

\appendix
\section{List of Coefficients}
\label{sec:A}

\subsection{Coefficients of the Accessory Parameter Expansions}
\label{sec:A.1}
The second-order coefficients of the accessory parameter expansions $E_{t}$ and $E_{q}$ are given below. For the expansion \eqref{eq:expansions_small}, valid in the regime $0 < t < q < 1 < \infty$, the coefficients are
\begin{align}
        &\epsilon_{1,1} = \frac{(1 + 4\delta_{1} - 4\delta_{\infty} - 4\rho^{2})(1 - 2\delta_{q} - 2\rho^{2} - 2\sigma^{2})(1 - 4\delta_{0} + 4\delta_{t} - 4\sigma^{2})}{8(1 - 4\rho^{2})(1 - 4\sigma^{2})}\,, \nonumber\\
        &\epsilon_{0,2} = 0\,, \nonumber\\
        &\epsilon_{2,0} = -\frac{1}{32}(1+14\delta_{0}-18\delta_{q}-18\delta_{t}-14\rho^{2}+13\sigma^{2}) - \frac{(\delta_{0}-\delta_{t})^{2}(1-4\delta_{q}-4\rho^{2})^{2}(1+12\sigma^{2})}{8(1-4\sigma^{2})^{3}} \nonumber\\
        &+\frac{(\delta_{q}^{2} - \rho^{2} + 2\delta_{q}\rho^{2} + \rho^{4})(16\delta_{0}^{2} + 8\delta_{0} - 32\delta_{0}\delta_{t} + 8\delta_{t} + 16\delta_{t}^{2} - 3)}{32(1-\sigma^{2})} + \frac{1}{32(1 - 4\sigma^{2})} \nonumber\\  &\times\Bigl[(1-4\delta_{q}-4\rho^{2})^{2}+4(\delta_{0}+\delta_{t})(1+16\rho^{2}-16(\delta_{q}+\rho^{2})^{2})-32(\delta_{0}+\delta_{t}-4\delta_{0}\delta_{t})(\delta_{q}^{2}+\rho^{4}) \nonumber\\
        &\qquad +14\delta_{0}-48\delta_{0}\delta_{q}-18\delta_{t}-8\delta_{0}\delta_{t}+80\delta_{q}\delta_{t} - 48\delta_{0}\rho^{2} - 64\delta_{0}\delta_{q}\rho^{2}+80\delta_{t}\rho^{2} - 128\delta_{0}\delta_{t}\rho^{2} \nonumber\\
        &\qquad- 64\delta_{q}\delta_{t}\rho^{2} + 256\delta_{0}\delta_{t}\delta_{q}\rho^{2}\Bigr]\,,
\end{align}
\begin{align}
        &\varepsilon_{1,1} = 0\,, \nonumber\\
        &\varepsilon_{2,0} = -\epsilon_{2,0}\,, \nonumber\\
        &\varepsilon_{0,2} = - \frac{1}{32}(1+14\delta_{\infty} - 18\delta_{1} - 18\delta_{q} - 14\sigma^{2}+13\rho^{2}) - \frac{(\delta_{1}-\delta_{\infty})^{2}(1-4\delta_{q}-4\sigma^{2})^{2}(1+12\rho^{2})}{8(1-4\rho^{2})^{3}} \nonumber\\
        &+\frac{(\delta_{q}^{2} - \sigma^{2} + 2\delta_{q}\sigma^{2} + \sigma^{4})(16\delta_{\infty}^{2} + 8\delta_{\infty} - 32\delta_{\infty}\delta_{1} + 8\delta_{1} + 16\delta_{1}^{2} - 3)}{32(1-\rho^{2})} + \frac{1}{32(1 - 4\rho^{2})} \nonumber\\
        &\times\Bigl[(1-4\delta_{q}-4\sigma^{2})^{2}+4(\delta_{1}+\delta_{\infty})(1+16\sigma^{2}-16(\delta_{q}+\sigma^{2})^{2})-32(\delta_{\infty}+\delta_{1} - 4\delta_{\infty}\delta_{1})(\delta_{q}^{2}+\sigma^{4}) \nonumber\\
        &\qquad +14\delta_{\infty}-48\delta_{\infty}\delta_{q}-18\delta_{1}-8\delta_{\infty}\delta_{1}+80\delta_{q}\delta_{1} - 48\delta_{\infty}\sigma^{2} - 64\delta_{\infty}\delta_{q}\sigma^{2} - 64\delta_{q}\delta_{1}\sigma^{2} +80\delta_{1}\sigma^{2} \nonumber\\
        &\qquad - 128\delta_{\infty}\delta_{1}\sigma^{2} + 256\delta_{\infty}\delta_{1}\delta_{q}\sigma^{2}\Bigr]\,.
\end{align}
whereas for the expasion \eqref{eq:expansions_large}, valid in the regime $0 < t < 1 < q < \infty$, they are
\begin{align}
        &\tilde{\epsilon}_{1,1} = \frac{(1-4\delta_{\infty}+\delta_{q}-4\rho^{2})(1-2\delta_{1}-2\rho^{2}-2\sigma^{2})(1-4\delta_{0}+4\delta_{t}-4\sigma^{2})}{8(1-4\rho^{2})(1-4\sigma^{2})}\,, \nonumber\\
        &\tilde{\epsilon}_{0,2} = 0\,, \nonumber\\
        &\tilde{\epsilon}_{2,0} = -\frac{1}{32}(1+14\delta_{0}-18\delta_{t}-18\delta_{1}-14\rho^{2}+13\sigma^{2})-\frac{(\delta_{0}-\delta_{t})^{2}(1-4\delta_{1}-4\rho^{2})^{2}(1+12\sigma^{2})}{8(1-4\sigma^{2})^{3}} \nonumber\\
        &+\frac{((\delta_{1}+\rho^{2})^{2}-\rho^{2})((4\delta_{0}-4\delta_{t}-1)^{2}+16\delta_{0}-4)}{32(1-\sigma^{2})}+\frac{1}{32(1-4\sigma^{2})}\Bigl[(1-4\rho^{2})^{2} \nonumber\\
        &+ 4(1+16\rho^{2}-16\rho^{4})(\delta_{0}-\delta_{t})^{2} - 64\delta_{1}(\delta_{0}^{2}+\delta_{t}^{2})(\delta_{1}+2\rho^{2}) - 32\delta_{1}(\delta_{0}+\delta_{t} - 4\delta_{0}\delta_{t})(\delta_{1}+2\rho^{2}) \nonumber\\
        &- 8\delta_{1}(1-2\delta_{1}-4\rho^{2}) - 2(1-4\rho^{2})^{2}(\delta_{0}+\delta_{t} + 16(1-4\rho^{2})(\delta_{0}-\delta_{t}) - 16\delta_{1}(3\delta_{0}-5\delta_{t})\Bigr]\,,
\end{align}
\begin{align}
        &\tilde{\varepsilon}_{1,1} = -\tilde{\epsilon}_{1,1}\,, \nonumber\\ 
        &\tilde{\varepsilon}_{2,0} = 0\,, \nonumber\\
        &\tilde{\varepsilon}_{0,2} = \frac{1}{32}(1-18\delta_{1}-18\delta_{q}+14\delta_{\infty}-14\sigma^{2}+13\rho^{2}) + \frac{(\delta_{\infty}-\delta_{q})^{2}(1-4\delta_{1}-4\sigma^{2})^{2}(1+12\rho^{2})}{8(1-4\rho^{2})^{3}} \nonumber\\
        &-\frac{((\delta_{1}+\sigma^{2})^{2}-\sigma^{2})((4\delta_{q}-4\delta_{\infty}-1)^{2}+16\delta_{q}-4)}{32(1-\rho^{2})}-\frac{1}{32(1-4\rho^{2})}\Bigl[(1-4\sigma^{2})^{2} \nonumber\\ 
        &+4(1+16\sigma^{2}-16\sigma^{4})(\delta_{\infty}-\delta_{q})^{2} - 64\delta_{1}(\delta_{q}^{2}+\delta_{\infty}^{2})(\delta_{1}+2\sigma^{2}) - 32\delta_{1}(\delta_{\infty}+\delta_{q} - 4\delta_{\infty}\delta_{q})(\delta_{1}+2\sigma^{2}) \nonumber\\
        &- 8\delta_{1}(1-2\delta_{1}-4\sigma^{2}) - 2(1-4\sigma^{2})^{2}(\delta_{\infty}+\delta_{q} + 16(1-4\sigma^{2})(\delta_{\infty}-\delta_{q}) - 16\delta_{1}(3\delta_{\infty} - 5\delta_{q})\Bigr]\,.
\end{align}
Further coefficients are available upon request.

\subsection{Coefficients of the separation constants}
\label{sec:A.2}
\begin{align}\label{eq:coeffs_betas}
    &\beta_{1,2,0,2} = \frac{1}{4}\biggl[\omega^{2} + 3m_{1}^{2} + m_{2}^{2} + m_{3}^{2} + 3\mu^{2} + 8 -4\sigma^{2} - 4\rho^{2} - \frac{(1 - m_{3}^{2})(\omega^{2} - (9+\mu^{2}))}{1-4\rho^{2}} \nonumber\\
    &\quad + \frac{(1 - 4\sigma^{2})(\omega^{2} - (9+\mu^{2}))}{1-4\rho^{2}} - \frac{(m_{1}^{2} - m_{2}^{2})(m_{3}^{2} - 4\rho^{2})}{1-4\sigma^{2}} - \frac{(m_{1}^{2} - m_{2}^{2})(\omega^{2} - (9 + \mu^{2}))}{1-4\rho^{2}} \nonumber\\
    &\quad - \frac{(m_{1}^{2}-m_{2}^{2})(m_{3}^{2}-4\rho^{2})(\omega^{2} - (9+\mu^{2}))}{(1-4\rho^{2})(1-4\sigma^{2})}\biggr]\,, \\
    &\beta_{1,0,2,2} = \frac{1}{4}\biggl[\omega^{2} + m_{1}^{2} + 3m_{2}^{2} + m_{3}^{2} + 3\mu^{2} + 8 -4\sigma^{2} - 4\rho^{2} - \frac{(1 - m_{3}^{2})(\omega^{2} - (9+\mu^{2}))}{1-4\rho^{2}} \nonumber\\
    &\quad + \frac{(1 - 4\sigma^{2})(\omega^{2} - (9+\mu^{2}))}{1-4\rho^{2}} + \frac{(m_{1}^{2} - m_{2}^{2})(m_{3}^{2} - 4\rho^{2})}{1-4\sigma^{2}} + \frac{(m_{1}^{2} - m_{2}^{2})(\omega^{2} - (9 + \mu^{2}))}{1-4\rho^{2}} \nonumber\\
    &\quad + \frac{(m_{1}^{2}-m_{2}^{2})(m_{3}^{2}-4\rho^{2})(\omega^{2} - (9+\mu^{2}))}{(1-4\rho^{2})(1-4\sigma^{2})}\biggr]\,, \\
    &\beta_{1,0,0,4} = \frac{1}{16}\left(8+m_{3}^{2}+\mu^{2} + 4\sigma^{2} - 6\rho^{2} + \omega^{2}\right)- \frac{(\omega^{2} + \mu^{2}+9)(m_{3}^{2}+4\sigma^{2})}{8(1-4\rho^{2})} \nonumber\\
    &\quad + \frac{((\omega^{2}-\mu^{2}-8)^{2}-4\omega^{2}-1)(4\sigma^{2}-m_{3}^{2}-1)^{2}}{32(1-4\rho^{2})} - \frac{((\omega^{2}-\mu^{2}-8)^{2}-4\omega^{2}-1)(1+4m_{3}^{2})}{32(1-4\rho^{2})} \nonumber\\
    &\quad - \frac{((4\sigma^{2} - m_{3}^{2}-1)^{2}-4m_{3}^{2})((\omega^{2}-\mu^{2}-8)^{2}-4\omega^{2})}{128(1-\rho^{2})}\nonumber\\
    &\quad + \frac{(1+12\rho^{2})(m_{3}^{2}-4\sigma^{2})(\omega^{2} - (9+\mu^{2}))^{2}}{32(1-4\rho^{2})^{3}}\,,
\end{align}
\begin{align}
    &\beta_{2,2,0,2} = \frac{1}{2}\left[(1 - 4\sigma^{2}) - (1 - 4\rho^{2}) + (m_{1}^{2} + m_{2}^{2} + m_{3}^{2} - 3) - \frac{(m_{1}^{2}-m_{2}^{2})(m_{3}^{2} - 4\rho^{2})}{1-4\sigma^{2}}\right]\,,\\
    &\beta_{2,0,2,2} = \frac{1}{2}\left[(1 - 4\sigma^{2}) - (1 - 4\rho^{2}) + (m_{1}^{2} + m_{2}^{2} + m_{3}^{2} - 3) + \frac{(m_{1}^{2}-m_{2}^{2})(m_{3}^{2}-4\rho^{2})}{1-4\sigma^{2}}\right]\,.
\end{align}

\section{Angular eigenvalues in the case: \texorpdfstring{$0 < t < 1 < q < \infty$}{0 < t < 1 < q < infinity}}
\label{sec:B}

Consider the following M\"{o}bius transformation
\begin{equation}\label{eq:varsigma}
    \varsigma = \frac{1 - \alpha_{2}^{2}}{1 - v^{2}}\,,
\end{equation}
which maps the singularities
\begin{equation}
        \left(\alpha_{1}^{2}, \alpha_{2}^{2}, \alpha_{3}^{2}, 1, \infty \right) \, \longmapsto \,   \left( t, 1, q, \infty, 0 \right)\,,
\end{equation}
with
\begin{equation}\label{eq:moduli_large}
    t = \frac{1 - \alpha_{2}^{2}}{1 - \alpha_{1}^{2}} \,, \qquad q = \frac{1 - \alpha_{2}^{2}}{1 - \alpha_{3}^{2}}\,.
\end{equation}
It turns out that Eq.~\eqref{eq:varsigma} results particularly convenient to study the limits $t \to 0$ and $q \to \infty$, which correspond to $\alpha_{2}$ and $\alpha_{3}$ close to one for fixed $\alpha_{1} \ll 1$. Furthermore, the $s$-homotopic transformation of the form
\begin{equation}
    Y(\varsigma) = \varsigma^{-1/2}(\varsigma - t)^{-1/2}(\varsigma - q)^{-1/2}(\varsigma - 1)^{-1/2}\mathcal{Y}(\varsigma)
\end{equation}
leads to a second order Fuchsian equation for $\mathcal{Y}(\varsigma)$ in the normal form
\begin{subequations}
    \begin{equation}\label{eq:fuchsian_varsigma}
	   \begin{split}
		\frac{d^{2}\mathcal{Y}}{d\varsigma^{2}} + \biggl[\frac{\delta_{0}}{\varsigma^{2}} &+ \frac{\delta_{t}}{(\varsigma - t)^{2}} + \frac{\delta_{1}}{(\varsigma - 1)^{2}} + \frac{\delta_{q}}{(\varsigma - q)^{2}} + \frac{\delta_{\infty} - \delta_{0} - \delta_{t} - \delta_{1} - \delta_{q}}{\varsigma(\varsigma - 1)}\\
		&\qquad\qquad + \frac{(t-1)E_{t}}{\varsigma(\varsigma - t)(\varsigma - 1)} + \frac{(q-1)E_{q}}{\varsigma(\varsigma - 1)(\varsigma - q)}\biggr]\mathcal{Y}(\varsigma) = 0,
	   \end{split}
    \end{equation}
where
    \begin{equation}\label{eq:delta_large}
        \begin{gathered}
	        \delta_{0} = \frac{1}{4}\left(1 - (9 + \mu^{2}) \right)\,, \quad \delta_{t} = \frac{1}{4}\left(1 - m_{1}^{2}\right)\,, \quad  \delta_{1} = \frac{1}{4}\left(1 - m_{2}^{2}\right)\,, \quad \delta_{q} = \frac{1}{4}\left(1 - m_{3}^{2}\right)\,,\\
            \delta_{\infty} = \frac{1}{4}\left(1 - \omega^{2}\right)\,,
        \end{gathered}
    \end{equation}
    \begin{equation}\label{eq:Et_large}
        \begin{split}
            E_{t} &=\frac{m_{1}^{2}}{4\alpha_{1}^{2}} + \frac{m_{1}^{2}}{4} - \frac{m_{1}\omega}{2\alpha_{1}} + \frac{(1 - \alpha_{1}^{2})}{2\alpha_{1}}\left[\frac{\alpha_{2}m_{2}m_{1}}{\alpha_{2}^{2} - \alpha_{1}^{2}} + \frac{\alpha_{3}m_{3}m_{1}}{\alpha_{3}^{2} - \alpha_{1}^{2}}\right] \\
            &- \frac{\left(\alpha_{1}^{4}\beta_{1} - \alpha_{1}^{2}\beta_{2} - \alpha_{1}^{6}\mu^{2} + \frac{\varPsi_{3}^{2}}{\chi_{3}}\right)}{4\alpha_{1}^{2}(\alpha_{2}^{2} - \alpha_{1}^{2})(\alpha_{3}^{2} - \alpha_{1}^{2})} - \frac{1}{2}\left(\frac{t}{t-1} + \frac{t}{t - q} - 2\right)\,,
        \end{split}
    \end{equation}
    \begin{equation}\label{eq:Eq_large}
        \begin{split}
            E_{q} &= \frac{m_{3}^{2}}{4\alpha_{3}^{2}} + \frac{m_{3}^{2}}{4} - \frac{m_{3}\omega}{2\alpha_{3}} + \frac{(1 - \alpha_{3}^{2})}{2\alpha_{3}}\left[\frac{\alpha_{1}m_{1}m_{3}}{\alpha_{3}^{2} - \alpha_{1}^{2}} + \frac{\alpha_{2}m_{2}m_{3}}{\alpha_{3}^{2} - \alpha_{2}^{2}}\right] \\
            &- \frac{\left(\alpha_{3}^{4}\beta_{1} - \alpha_{3}^{2}\beta_{2} - \alpha_{3}^{6}\mu^{2} + \frac{\varPsi_{3}^{2}}{\chi_{3}}\right)}{4\alpha_{3}^{2}(\alpha_{3}^{2} - \alpha_{1}^{2})(\alpha_{3}^{2} - \alpha_{2}^{2})} - \frac{1}{2}\left(\frac{q}{q-1} + \frac{q}{q - t} - 2\right)\,.
        \end{split}
    \end{equation}
\end{subequations}
We will consider the accessory parameter expansions in the small $t$ and large $q$ limit \eqref{eq:ape_large} and \eqref{eq:moduli_large}. Thus, $\alpha_{2}$ and $\alpha_{3}$ can be expressed in terms of $t$ and $q$ as
\begin{equation}
    \alpha_{2} = \sqrt{1 - t(1-\alpha_{1}^{2})}\,, \qquad \alpha_{3} = \sqrt{1-\frac{t(1-\alpha_{1}^{2})}{q}}
\end{equation}
where $0 < t < 1 < q$ and $0 <\alpha_{1} < 1$ imply $\alpha_{2}$ and $\alpha_{3}$ close to one, satisfying the hierarchy between the rotation parameters $0 < \alpha_{1} < \alpha_{2} < \alpha_{3}$.

The accessory parameter expansions of $E_{t}$ and $E_{q}$ given by \eqref{eq:Et_ape_large} and \eqref{eq:Eq_ape_large} are equal to the accessory parameters of the angular ODE \eqref{eq:Et_large} and \eqref{eq:Eq_large}, respectively.
Therefore, the perturbative expansion of the separation constants are
\begin{subequations}\label{eq:beta_large_ser}
    \begin{equation}
        \begin{split}
            \beta_{1} = 3 &+ (1-4\sigma^{2}) + m_{1}^{2} + (m_{2}+m_{3}-\omega)^{2} + 2\mu^{2} + 2m_{1}\left(m_{2}+m_{3}-\omega\right)\alpha_{1}\\
            &+\frac{1}{2}\Biggl[2m_{2}(\omega-m_{3})-\left(m_{1}^{2}+m_{2}^{2}+\mu^{2}-(1-4\rho^{2})+(1-4\sigma^{2})\right)\\&+ \frac{(m_{1}^{2}-(9+\mu^{2}))(m_{2}^{2}-1)}{(1-4\sigma^{2})} + \frac{(m_{1}^{2} - (9+\mu^{2}))(1-4\rho^{2})}{(1-4\sigma^{2})}\Biggr]t + \ldots
        \end{split}
    \end{equation}
    \begin{equation}
        \begin{split}
            \beta_{2} = 3 &+ (1-4\sigma^{2}) + \mu^{2} + 2m_{1}^{2} + (m_{2}+m_{3}-\omega)^{2} + 4m_{1}(m_{2}+m_{3}-\omega)\alpha_{1}\\
            &- \frac{1}{2}\Biggl[(1-4\sigma^{2}) + (1-4\rho^{2}) + 3m_{1}^{2} + \mu^{2} +m_{2}^{2} + 2m_{2}(m_{3}-\omega) + 2(m_{3}-\omega)^{2}\\ 
            &-\frac{(m_{1}^{2}-(9+\mu^{2}))(m_{2}^{2}-1)}{(1-4\sigma^{2})}-\frac{(m_{1}^{2}-(9+\mu^{2}))(1-4\rho^{2})}{(1-4\sigma^{2})}\Biggr]t + \ldots 
        \end{split}
    \end{equation}
\end{subequations}

\section{Radial eigenfrequencies in the case: \texorpdfstring{$0 < t < q < 1$}{0 < t < q < 1}}
\label{sec:C}

Under the M\"{o}bius transformation
\begin{equation}\label{eq:Mobius_small}
    z = \frac{u^{2} - u_{1}^{2}}{u^{2} - u_{0}^{2}}\,,
\end{equation}
the singular points of the radial ODE~\eqref{eq:dim_radial_ode} are mapped as follows
\begin{subequations}
\begin{equation}\label{eq:z_to_u_small}
    \left( u_{0}^{2}, u_{1}^{2}, u_{2}^{2}, u_{3}^{2}, \infty \right) \,\longmapsto \, \left( \infty, 0, t, q, 1\right)\,,
\end{equation}
with
\begin{equation}\label{eq:moduli_rad_small}
    t = \frac{u_{2}^{2} - u_{1}^{2}}{u_{2}^{2} - u_{0}^{2}}, \qquad q = \frac{u_{3}^{2} - u_{1}^{2}}{u_{3}^{2} - u_{0}^{2}}\,.
\end{equation}
\end{subequations}
so that for $u_{0}^{2} <u_{1}^{2} < 0 < u_{2}^{2} < u_{3}^{2}$, we have $0 < t < q < 1$. We next perform the following $s$-homotopic transformation:
\begin{equation}
    R(z) = z^{-1/2}(z-t)^{-1/2}(z-q)^{-1/2}(z-1)\psi(z)\,,
\end{equation}
under which the differential equation for $\psi(z)$ takes the form
\begin{equation}\label{eq:fuchsian_rad_small}
	\begin{split}
		\frac{d^{2}\mathcal{\psi}}{dz^{2}} + \biggl[\frac{\delta_{0}}{z^{2}} &+ \frac{\delta_{t}}{(z - t)^{2}} + \frac{\delta_{q}}{(z - q)^{2}} + \frac{\delta_{1}}{(z - 1)^{2}}  + \frac{\delta_{\infty} - \delta_{0} - \delta_{t} - \delta_{1} - \delta_{q}}{z(z - 1)}\\
    	&\qquad\qquad + \frac{(t-1)E_{t}}{z(z - t)(z - 1)} + \frac{(q-1)E_{q}}{z(z - q)(z - 1)}\biggr]\psi(z) = 0,
	\end{split}
\end{equation}
where
\begin{equation}\label{eq:delta_radial_small}
    \begin{gathered}
	    \delta_{0} = \frac{1}{4}\left(1 - \theta_{u_{1}}^{2}\right)\,, \quad \delta_{t} = \frac{1}{4}\left(1 - \theta_{u_{2}}^{2}\right)\,, \quad \delta_{q} = \frac{1}{4}\left(1 - \theta_{u_{3}}^{2}\right)\,, \\ 
        \delta_{1} = \frac{1}{4}\left(1 - (9 + \mu^{2}) \right)\,, \quad \delta_{\infty} = \frac{1}{4}\left(1 - \theta_{u_{0}}^{2}\right)\,.
    \end{gathered}
\end{equation}
The accessory parameters are given by
\begin{subequations}\label{eq:accessory_small}
    \begin{align}
        E_{t} &= \frac{\left(u_{2}^{2}\beta_{1} + \beta_{2} + u_{2}^{4}\mu^{2} + \frac{\varPsi_{3}^{2}}{u_{2}^{2}\chi_{3}} \right)}{4(u_{0}^{2} - u_{1}^{2})(u_{2}^{2} - u_{3}^{2})} - \frac{\left(u_{2}^{2} - u_{0}^{2}\right)\left(u_{2}^{2} - u_{1}^{2}\right)}{2u_{2}(u_{0}^{2} - u_{1}^{2})}A_{2}
        +\frac{\left(u_{2}^{2} - u_{1}^{2}\right)}{4\left(u_{0}^{2} - u_{1}^{2}\right)}\theta_{u_{2}}^{2} \label{eq:Et_radial_small}\\ 
        &- \frac{\left(u_{1}^{2} - u_{2}^{2}\right)\left(u_{0}^{2} + u_{2}^{2}\right)}{8u_{2}^{2}\left(u_{0}^{2} - u_{1}^{2}\right)}\theta_{u_{2}}^{2} + \frac{\left(u_{0}^{2} - u_{2}^{2}\right)\left(u_{1}^{2} - u_{2}^{2}\right)}{2\left(u_{0}^{2} - u_{1}^{2}\right)\left(u_{2}^{2} - u_{3}^{2}\right)} - \frac{\left(u_{0}^{2} - u_{2}^{2}\right)}{2\left(u_{0}^{2} - u_{1}^{2}\right)}\,, \notag\\
        E_{q} &= \frac{\left(u_{3}^{2}\beta_{1} + \beta_{2} + u_{3}^{4}\mu^{2} + \tfrac{\varPsi_{3}^{2}}{u_{3}^{2}\chi_{3}}\right)}{4(u_{0}^{2} - u_{1}^{2})(u_{3}^{2} - u_{2}^{2})} - \frac{\left(u_{3}^{2} - u_{0}^{2}\right)\left(u_{3}^{2} - u_{1}^{2}\right)}{2u_{3}(u_{0}^{2} - u_{1}^{2})}A_{3} +\frac{\left(u_{3}^{2} - u_{1}^{2}\right)}{4\left(u_{0}^{2} - u_{1}^{2}\right)}\theta_{u_{3}}^{2} \label{eq:Eq_radial_small}\\
        &- \frac{\left(u_{1}^{2} - u_{3}^{2}\right)\left(u_{0}^{2} + u_{3}^{2}\right)}{8u_{3}^{2}\left(u_{0}^{2} - u_{1}^{2}\right)}\theta_{u_{3}}^{2} + \frac{\left(u_{3}^{2} - u_{1}^{2}\right)\left(u_{3}^{2} - u_{0}^{2}\right)}{2\left(u_{0}^{2} - u_{1}^{2}\right)\left(u_{3}^{2} - u_{2}^{2}\right)}-\frac{\left(u_{3}^{2} - u_{0}^{2}\right)}{2\left(u_{0}^{2} - u_{1}^{2}\right)} \notag\,,
    \end{align}
\end{subequations}
where $A_{2}$ and $A_{3}$ are determined from Eq.~\eqref{eq:coeffs_rad}, whereas $\theta_{u_{2}}$ and $\theta_{u_{3}}$ are defined by Eq.~\eqref{eq:thetas}, and $\varPsi_{3}$ and $\chi_{3}$ are defined by Eq.~\eqref{eq:varPsis}.

In the regime $0 < t < q < 1 $, we make the ansatz 
\begin{equation}\label{eq:sigma_and_rho_small}
    \sigma = \sum_{i,j = 0}^{\infty}\varsigma_{i,j}\left(\frac{t}{q}\right)^{i}q^{j}\,, \qquad \rho = \sum_{i,j = 0}^{\infty}\varrho_{i,j}\left(\frac{t}{q}\right)^{i}q^{j}\,,
\end{equation}
and determine the coefficients recursively by inverting the accessory parameter expansions Eqs.~\eqref{eq:Et_ape_small} and \eqref{eq:Eq_ape_small}, solving order by order for the coefficients in terms of the parameters $\delta_{i}$, $E_{t}$, and $E_{q}$. The leading terms of these double expansions are
\begin{subequations}
    \begin{equation}\label{eq:sigma_expansion}
            \sigma = \frac{1}{2}\sqrt{1-4E_{t}-4\delta_{0}-4\delta_{t}} + \frac{E_{q}(E_{t}+2\delta_{t})}{2(E_{t}+\delta_{0}+\delta_{t})\sqrt{1-4E_{t}-4\delta_{0}-4\delta_{t}}}\frac{t}{q} + \mathcal{O}\left(t^{2}q^{-2},t,q^{2}\right)\,,
    \end{equation}
    \begin{equation}\label{eq:rho_expansion}
        \begin{split}
            &\rho = \frac{1}{2}\sqrt{1-4E_{q}-4E_{t}-4\delta_{0}-4\delta_{t}-4\delta_{q}} \\
            &\qquad +\frac{(E_{q}+2\delta_{q})(E_{q}+E_{t}+\delta_{0}+\delta_{t}+\delta_{q}+\delta_{1}-\delta_{\infty})}{2(E_{q}+E_{t}+\delta_{0}+\delta_{t}+\delta_{q})\sqrt{1-4E_{q}-4E_{t}-4\delta_{0}-4\delta_{t}-4\delta_{q}}}q + \mathcal{O}\left(t^{2}q^{-2},t,q^{2}\right)\,,
        \end{split}
    \end{equation}
\end{subequations}
where $\varsigma_{i,j} = 0$ for $i < j$ and $\varrho_{i,j} = 0$
for $i > j$, as discussed in \cite{Liu:2024eut}. As presented in Section~\ref{sec:3.2}, the computation of $\sigma$ is affected by the mixing of perturbative orders encountered in the small $t$, large $q$ regime. Nevertheless, the expansion of $\rho$ does not exhibit this behavior, so that substituting Eqs.~\eqref{eq:moduli_rad_small}, \eqref{eq:delta_radial_small}, and \eqref{eq:Eq_radial_small} into the Floquet exponent expansion~\eqref{eq:rho_expansion} and expanding perturbatively in the horizon radius and the rotation parameters yields
\begin{equation}\label{eq:rho_rad_small}
    \begin{split}
        \rho &= \frac{\ell+2}{2} + \Biggl[\frac{15\ell(\ell+2)(\ell+4)}{32(\ell+1)(\ell+3)} - \frac{\mu^{2}(\mu^{2}-6\omega^{2}+6\ell(\ell+4)+32)}{32(\ell+1)(\ell+2)(\ell+3)}\\
        &\qquad - \frac{5\omega^{2}(\omega^{2} - 6\ell(\ell+4)-28)}{32(\ell+1)(\ell+2)(\ell+3)}\Biggr]u_{3}^{4} + \ldots\,,
    \end{split}
\end{equation}
which coincides with Eq.~\eqref{eq:rho_rad}. Furthermore, the radial quantization condition that reproduces the real part of the QNM frequencies is given by
\begin{equation}\label{eq:qnm_cond_small}
    \frac12 + \rho + a_{1} \pm a_{\infty} = -n\,, \qquad n \in \mathbb{Z}_{\geq 0}\,.
\end{equation}

\bibliography{References}
\bibliographystyle{JHEP}

\end{document}